\documentclass[aps, prd, preprint, amsfonts,
  amssymb, amsmath, showpacs, a4paper, nofootinbib]{revtex4-1}

\usepackage{braket,slashed}

\usepackage{color,graphicx}

\usepackage{subfig}

\usepackage{soul}

\begin{document}


\title{Fifth forces, Higgs portals and broken scale invariance}

\author{Clare Burrage}
\email{clare.burrage@nottingham.ac.uk}
\affiliation{School of Physics and Astronomy, University of Nottingham,\\ Nottingham NG7 2RD, United Kingdom}

\author{Edmund J. Copeland}
\email{edmund.copeland@nottingham.ac.uk}
\affiliation{School of Physics and Astronomy, University of Nottingham,\\ Nottingham NG7 2RD, United Kingdom}

\author{Peter Millington}
\email{p.millington@nottingham.ac.uk}
\affiliation{School of Physics and Astronomy, University of Nottingham,\\ Nottingham NG7 2RD, United Kingdom}

\author{Michael Spannowsky}
\email{michael.spannowsky@durham.ac.uk}
\affiliation{Institute for Particle Physics Phenomenology, Department of Physics, Durham University,\\ Durham DH1 3LE, United Kingdom}


\preprint{IPPP/18/23}



\begin{abstract}
We study the relationship between the strength of fifth forces and the origin of scale breaking in the Standard Model (SM) of particle physics. We start with a light scalar field that is conformally coupled to a toy SM matter sector through a Weyl rescaling of the metric.  After appropriately normalizing the fields, the conformally coupled scalar only interacts directly with the would-be Higgs field through kinetic-mixing and Higgs-portal terms. Thus, for the first time, we describe the equivalence of conformally coupled scalar-tensor modifications of gravity and Higgs-portal theories, and we find that the usual tree-level fifth forces only emerge if there is mass mixing between the conformally coupled scalar and the Higgs field. The strength of the fifth force, mediated by the light scalar, then depends on whether the mass of the Higgs arises from an explicit symmetry-breaking term or a spontaneous mechanism of scale breaking. Solar System tests of gravity and the non-observation of fifth forces therefore have the potential to provide information about the structure of the Higgs sector and the origin of its symmetry breaking, setting an upper bound on the magnitude of any explicit scale-breaking terms. These results demonstrate the phenomenological importance (both for cosmology and high-energy physics) of considering how scalar-tensor modifications of gravity are embedded within extensions of the SM.
\end{abstract}
 
\maketitle


\section{Introduction}

The desire to accommodate both early- and late-time accelerated expansion within minimal extensions of the Standard Model (SM) of particle physics has motivated renewed interest in theories involving non-minimal couplings to the scalar curvature: so-called scalar-tensor theories. If any of the non-minimally coupled degrees of freedom are light, they are able to mediate long-range forces,  and such new physics at low energy scales could be connected to possible solutions of the cosmological constant problem~\cite{Copeland:2006wr,Clifton:2011jh,Bull:2015stt,Joyce:2014kja}.  Many attempts have been made to search for light scalar fields mediating long-range fifth forces \cite{Adelberger:2003zx}, so far without success, and it remains an open question whether such weakly coupled and light new physics is allowed to exist in our universe.

Scalar-tensor theories fall into the broad class of models known as modified gravity.  The recent observation of the gravitational-wave signal from a neutron-star merger by LIGO-Virgo and an associated electromagnetic event, in particular by the gamma-ray satellites Fermi and INTEGRAL~\cite{TheLIGOScientific:2017qsa,GBM:2017lvd}, showed that gravitational waves travel at the same speed as photons to around $1$ part in $10^{15}$. This result has led to a class of modified gravity models being excluded as the sole explanation of dark energy~\cite{Creminelli:2017sry,Sakstein:2017xjx,Ezquiaga:2017ekz,Baker:2017hug,Crisostomi:2017lbg,Langlois:2017dyl,Dima:2017pwp}. However, the conformally coupled scalar-tensor theories that we study in this work do not fall into this class and always predict equal speeds of propagation for gravitational and electromagnetic waves.

However, in order to be compatible with Solar System tests of gravity, it is widely assumed that  one must either fine-tune the couplings to matter or introduce some dynamic mechanism of screening in order to hide the associated scalar fifth forces from these local observations. The latter provides a serious phenomenological challenge for modified theories of gravity, and it continues to attract significant attention (for a review see, e.g., Ref.~\cite{Clifton:2011jh}). However, it has also been argued that fifth-force constraints can be evaded entirely if the SM extension and its coupling(s) to gravity are scale (or globally Weyl) invariant~\cite{Brax:2014baa,Ferreira:2016kxi}, since the conformal transformation to the Einstein frame yields at most derivative couplings of the additional scalar degree(s) of freedom to SM fields~\cite{Shaposhnikov:2008xb}. A particular example of such a theory is the Higgs-dilaton model studied in Refs.~\cite{Wetterich:1987fm,Buchmuller:1988cj,Shaposhnikov:2008xb,Shaposhnikov:2008xi,Blas:2011ac,GarciaBellido:2011de,GarciaBellido:2012zu,Bezrukov:2012hx,Henz:2013oxa,Rubio:2014wta,Karananas:2016grc,Ferreira:2016vsc,Ferreira:2016wem,Ferreira:2018qss}. In order to prevent the breaking of Weyl invariance by loop corrections~\cite{Coleman:1973jx}, infrared divergences must be regulated by introducing a mass scale that depends entirely on the dynamical fields~\cite{Englert:1976ep,Shaposhnikov:2008xi,Ghilencea:2015mza,Ghilencea:2016ckm,Ferreira:2018itt}. In this way, one can maintain Weyl symmetry at the loop level, albeit with the loss of renormalizability. Alternatively, one can exploit dimensional transmutation~\cite{Coleman:1973jx} and the associated loop-level breaking of scale symmetry to construct renormalizable theories of gravity~\cite{Salvio:2014soa,Einhorn:2014gfa, Einhorn:2015lzy, Einhorn:2016mws, Einhorn:2017icw}, to stabilize the Planck mass in scalar-tensor theories in the presence of sources of explicit scale breaking~\cite{Kannike:2016wuy} or to introduce radiatively-generated screening mechanisms~\cite{Burrage:2016xzz}.

Previous studies in the literature have focused on theories that are fully scale invariant. In this work, we show how the loss of scale invariance reintroduces scalar-mediated fifth forces and demonstrate how the strength of fifth forces depends on the amount of explicit scale breaking present in the SM.  We will see that the most important scale-breaking parameter in the SM is the mass of the Higgs (which may also be generated, in full or in part, by spontaneous symmetry breaking).  The Higgs mass has been precisely measured at the LHC to be $125.18\pm 0.16\ {\rm GeV}$~\cite{Tanabashi:2018oca}, but determining whether this mass arises from an explicit mass scale or from another symmetry-breaking mechanism is beyond the reach of current collider experiments. In this work, we consider both possibilities, including the case where the observed Higgs mass arises from a combination of explicit and spontaneously generated scales, and we show how this affects the strength of the fifth force mediated between both elementary fermions and hadronic matter by any light conformally coupled scalars in the theory.  Moreover, due to the fact that a large proportion of the coupling to hadronic matter is induced by the conformal anomaly, the strength of the fifth force between hadrons may be parametrically smaller than that occurring between elementary particles of the same mass, potentially giving rise to effective violations of the weak equivalence principle that warrant further study beyond this work.

Our results illustrate a new explanation for the non-observation of fifth forces mediated by the light scalars that are so common in theories of new physics: the suppression of explicit scale-breaking terms in the SM (viz.~the bare SM Higgs mass).  In non-minimally coupled scalar-tensor theories, the interaction between the new scalar degree of freedom $\chi$ and the Higgs boson of the SM results in a natural way in a Higgs-portal model \cite{Schabinger:2005ei, Patt:2006fw}, that is to say: non-minimally coupled scalar-tensor theories are equivalent to Higgs-portal theories. This opens up two avenues to tension such models with experimental data. On the one hand, it presents a novel opportunity to study Higgs physics with experiments more commonly considered tests of gravity. On the other hand, precision measurements of the Higgs boson's properties, e.g., its branching ratio into invisible final states \cite{Aaboud:2017bja, Khachatryan:2016whc} or its total width~\cite{Aad:2015xua, Sirunyan:2017exp} can be used to set indirect limits on the interactions with $\chi$ \cite{Djouadi:2011aa, Englert:2013gz}. As a consequence, Higgs phenomenology is directly impacted by any screening mechanism of $\chi$ \cite{inprep} and vice versa. For example, the radiative screening mechanism described in Ref.~\cite{Burrage:2016xzz} can be viewed as a light Higgs portal to a hidden Coleman-Weinberg sector (see, e.g., Refs.~\cite{Coleman:1973jx, Hempfling:1996ht, Meissner:2006zh, Chang:2007ki,Foot:2007as,Iso:2009ss,AlexanderNunneley:2010nw,Englert:2013gz}). Moreover, this observed equivalence means that models of dark matter involving singlet 	scalars, which communicate with the SM via Higgs portals~\cite{Silveira:1985rk,McDonald:1993ex,Burgess:2000yq,Davoudiasl:2004be,Patt:2006fw}, have much more in common with conformally coupled scalar-tensor theories than previously realized. In certain regions of parameter space, or at certain epochs in the cosmological evolution, this may open up new ways to exploit the well-studied phenomenology of certain modifications of gravity in the context of more traditional theories of dark matter.

The remainder of this article is organized as follows. In Sec.~\ref{sec:ccscalars}, we consider the Brans-Dicke theory~\cite{Brans:1961sx} and two prototypal models of screened fifth forces: the chameleon \cite{Khoury:2003aq,Khoury:2003rn,Khoury:2013yya} and symmetron~\cite{Hinterbichler:2010es,Hinterbichler:2011ca}, and illustrate how the coupling of the light scalar degree of freedom to fermions arises through kinetic and mass mixings with the would-be SM Higgs. The contributions of the resulting fifth forces to the Yukawa potential are calculated in Sec.~\ref{sec:fifthforce}, and we show that it is the mass mixing that provides the dominant source of long-range fifth forces. In Sec.~\ref{sec:ExplicitScale}, we consider the case in which the electroweak scale is generated through a combination of explicit and spontaneous scale breaking and, after extending the analysis to baryonic matter, we derive an upper bound on the magnitude of the explicit scale breaking in the Higgs sector from the non-observation of fifth forces. In Sec.~\ref{sec:HiggsDilaton}, we consider the case in which all dimensionful scales are generated spontaneously --- the so-called Higgs-dilaton model --- illustrating how the kinetic mixing is eliminated in the fully scale-invariant limit. Linear-order analyses of this model in both the Jordan and Einstein frames are provided in the appendices for completeness. Our concluding remarks are presented in Sec.~\ref{sec:Conclusions}.


\section{Conformally coupled scalars}
\label{sec:ccscalars}

When writing down conformally coupled\footnote{By ``conformally coupled’’, we are referring generally to any field with a non-minimal coupling to the Ricci scalar that can be removed by a Weyl rescaling of the metric.} scalar-tensor theories, we have to make a choice of frame. This choice of frame does not affect physical observables; it just changes whether the scalar field appears coupled explicitly to the scalar curvature --- the Jordan frame --- or whether gravity is described by the standard Einstein Hilbert term and the scalar field defines a rescaled metric on which matter particles move --- the Einstein frame. In the following sections, the Einstein frame will prove to be the most convenient for our calculations. However, in the context of the Higgs-dilaton model of Sec.~\ref{sec:HiggsDilaton}, we will show explicitly how the same results can be obtained in both frames (see the appendices).  

The actions of conformally coupled scalar-tensor theories can be written in the Einstein frame in the following generic form
\begin{equation}
\label{eq:startingaction}
S\ = \ \int\!{\rm d}^4x\,\sqrt{-\,\tilde{g}}\bigg[\frac{M_{\rm Pl}^2}{2}\,\tilde{\mathcal{R}}\:-\:\frac{1}{2}\,\tilde{g}^{\mu\nu}\,\partial_{\mu}\chi\,\partial_{\nu}\chi\:-\:V(\chi)\bigg]\:+\:S_{\rm SM}\big[A^2(\chi)\tilde{g}_{\mu\nu},\{\psi\}\big]\;,
\end{equation}
where we use a tilde to indicate the Einstein-frame metric $\tilde{g}_{\mu\nu}$. The SM degrees of freedom $\{\psi\}$ move on geodesics determined by the Jordan-frame metric $g_{\mu\nu}=A^2(\chi)\tilde{g}_{\mu\nu}$. Note that we have set to zero any bare Higgs-portal couplings between the conformally coupled scalar $\chi$ and the SM Higgs field in the Jordan frame. Throughout this article, we work with the ``mostly plus'' signature convention $(-,+,+,+)$.

Scalar tensor theories of the form in Eq.~\eqref{eq:startingaction} are expected to possess fifth forces mediated by the scalar $\chi$, without any further need to understand the structure of the SM. One aim of the present article is to show explicitly why this is not the case, and we begin by sketching the standard argument for the presence of fifth forces.

In the case that $A^2(\chi)$ can be expanded as $A^2(\chi)=1+\frac{\chi^n}{M^n}+\dots$, we have
\begin{equation}
S_{\rm SM}[A^2(\chi)\tilde{g}_{\mu\nu},\{\psi\}]\ =\ S_{\rm SM}[\tilde{g}_{\mu\nu},\{\psi\}]\:+\:\frac{\delta S_{\rm SM}[g_{\mu\nu},\{\psi\}]}{\delta g_{\mu\nu}}\bigg|_{g\,=\,\tilde{g}}\,\frac{\chi^n}{M^n}\,\tilde{g}_{\mu\nu}\:+\:\dots\;,
\end{equation}
where we are imagining that $n=2$ or $n=1$, depending on whether the coupling function $A^2(\chi)$ is $\mathbb{Z}_2$ symmetric or not. Since
\begin{equation}
\frac{\delta S_{\rm SM}[g_{\mu\nu},\{\psi\}]}{\delta g_{\mu\nu}}\ =\ \frac{1}{2}\,T^{\mu\nu}
\end{equation}
is the energy-momentum tensor of the matter fields, we find that
\begin{equation}
S_{\rm SM}[A^2(\chi)\tilde{g}_{\mu\nu},\{\psi\}]\ =\ S_{\rm SM}[\tilde{g}_{\mu\nu},\{\psi\}]\:+\:\frac{\chi^n}{2M^n}\,\tilde{g}_{\mu\nu}\tilde{T}^{\mu\nu}\:+\:\dots\;.
\end{equation}
Commonly, fifth forces are estimated by modeling the SM degrees of freedom by a pressureless perfect fluid, i.e.~taking $\tilde{T}=-\,\rho$ (where $\rho$ is the non-relativistic energy density of the matter fields). Doing so, we would conclude that the SM degrees of freedom, through the trace of their energy-momentum tensor, are coupled universally to the scalar $\chi$, experiencing a fifth force
\begin{equation}
\label{eq:usualfifth}
\vec{F}\ =\ -\,\frac{n}{2}\,\frac{\chi^{n-1}}{M^{n-1}}\,\vec{\nabla}\,\frac{\chi}{M}\;.
\end{equation}
whenever there is a spatially varying scalar field profile. Such scalar profiles are, for instance, sourced by the energy-momentum tensors of massive bodies, thereby giving rise to fifth forces on test particles in their vicinity and leading to stringent constraints on conformally coupled scalars from tests of general relativity.

To see why the above argument does not capture all of the relevant physics, we consider a toy model for the SM, written in terms of the Jordan-frame metric $g_{\mu\nu}$ as
\begin{align}
S_{\rm SM}[g_{\mu\nu},\{\psi\}]\ &=\ \int\!{\rm d}^4x\,\sqrt{-\,g}\bigg[-\:\frac{1}{2}\,g^{\mu\nu}\,\partial_{\mu}\phi\,\partial_{\nu}\phi\:+\:\frac{1}{2}\,\mu^2\,\phi^2\:-\:\frac{\lambda}{4!}\,\phi^4\:-\:\frac{3}{2}\,\frac{\mu^4}{\lambda}\nonumber\\&\qquad -\:\bar{\psi}ie_{a}^{\mu}\gamma^{a}\overset{\leftrightarrow}{\partial}_{\mu}\psi\:-\:y\,\bar{\psi}\phi\psi\bigg]\;,
\end{align}
where $\overset{\leftrightarrow}{\partial}_{\mu}\equiv \frac{1}{2}\big(\overset{\rightarrow}{\partial}_{\mu}-\overset{\leftarrow}{\partial}_{\mu}\big)$, allowing us to omit the spin connection from the action (see, e.g., Ref.~\cite{Ferreira:2016kxi}). The real scalar field $\phi$ plays the role of the SM Higgs, and $\psi$ describes a Dirac fermion, whose mass arises via its Yukawa coupling after the Higgs undergoes spontaneous symmetry breaking. The constant shift in the potential ensures a vanishing classical contribution to the Jordan-frame cosmological constant in the symmetry-broken phase. We have written the Dirac operator in terms of the vierbein $e^{\mu}_a$, where the indices $\mu$ and $a$ label the coordinates of the curved and Minkowski spaces, respectively, i.e.~$g_{\mu\nu}=\eta_{ab}\,e^{a}_{\mu}e^{b}_{\nu}$. Hereafter, for notational simplicity, we will simply write $e_{a}^{\mu}\gamma^a\partial_{\mu}\equiv \slashed{\partial}$ when appropriate to do so.

So as to make the scalar couplings to matter explicit, we proceed by rewriting the theory in terms of the  Einstein-frame metric $\tilde{g}_{\mu\nu}$. The action is then
\begin{align}
S_{\rm SM}[A^2(\chi)\tilde{g}_{\mu\nu},\{\psi\}]\ &=\ \int\!{\rm d}^4x\,\sqrt{-\,\tilde{g}}\bigg[-\:\frac{1}{2}\,A^2(\chi)\tilde{g}^{\mu\nu}\,\partial_{\mu}\phi\,\partial_{\nu}\phi\:+\:\frac{1}{2}\,A^4(\chi)\,\mu^2\,\phi^2\nonumber\\&\quad -\:\frac{\lambda}{4!}\,A^4(\chi)\,\phi^4\:-\:\frac{3}{2}\,A^4(\chi)\,\frac{\mu^4}{\lambda}\:-\:A^2(\chi)\,\bar{\psi}i\overset{\leftrightarrow}{\slashed{\partial}}\psi\:-\:y\,A^4(\chi)\,\bar{\psi}\phi\psi\bigg]\;.
\end{align}
After redefining the Higgs and fermion fields according to their scaling dimensions as
\begin{equation}
\label{eq:phiredef}
\tilde{\phi}\ \equiv\ A(\chi)\phi\;,\qquad \tilde{\psi}\ \equiv\ A^{3/2}(\chi)\psi\;,
\end{equation}
such that the system is as close to being canonically normalized as possible, our toy SM Lagrangian becomes
\begin{align}
\label{eq:LSMEframe}
\tilde{\mathcal{L}}\ &=\ -\:\frac{1}{2}\,\tilde{g}^{\mu\nu}\,\partial_{\mu}\tilde{\phi}\,\partial_{\nu}\tilde{\phi}\:+\:\tilde{g}^{\mu\nu}\,\tilde{\phi}\,\partial_{\mu}\tilde{\phi}\,\partial_{\nu}\ln A(\chi)\:-\:\frac{1}{2}\,\tilde{g}^{\mu\nu}\,\tilde{\phi}^2\,\partial_{\mu}\ln A(\chi)\,\partial_{\nu}\ln A(\chi)\nonumber\\&\qquad+\:\frac{1}{2}\,\mu^2\,A^2(\chi)\,\tilde{\phi}^2\:-\:\frac{\lambda}{4!}\,\tilde{\phi}^4\:-\:\frac{3}{2}\,A^4(\chi)\,\frac{\mu^4}{\lambda}\:-\:\bar{\tilde{\psi}}i\overset{\leftrightarrow}{\tilde{\slashed{\partial}}}\tilde{\psi}\:-\:y\,\bar{\tilde{\psi}}\tilde{\phi}\tilde{\psi}\;,
\end{align}
where $\tilde{\slashed{\partial}}\equiv \tilde{e}_{a}^{\mu}\gamma^a\partial_{\mu} = A^{-1}(\chi)e_{a}^{\mu}\gamma^a\partial_{\mu}$.

By inspection of Eq.~\eqref{eq:LSMEframe}, we see that the conformally coupled scalar $\chi$ does not couple directly to the fermions, and it couples to the Higgs field only derivatively and through the Higgs-portal term proportional to the bare mass of the Higgs field. The reason for this is that, with the exception of the Higgs kinetic and mass terms, the SM Lagrangian is locally Weyl invariant, and therefore invariant under conformal rescalings. However, we will show that the presence of the Higgs mass term is sufficient to give rise to long-range fifth forces between the fermions that are, in the SM, independent of the electroweak scale. Moreover, and as we describe in detail in what follows, the usual tree-level fifth force can only arise if there is a mass mixing between $\phi$ and $\chi$. This is impossible above the electroweak phase transition, since the Higgs-portal term $A^2(\chi)\tilde{\phi}^2$ is quadratic in the fluctuations of the Higgs field. This may be important for understanding the impact of these fields throughout the history of the universe. In particular, we see that the behaviour of these modified gravity theories is significantly different before and after the electroweak phase transition.  Below the electroweak phase transition, when the Higgs obtains a non-zero vev $v_{\phi}$, the necessary mass mixing can arise if $A^2(\chi)$ is linear in the $\chi$ fluctuations, and this linear dependence on the field fluctuations can be realized only if there is an explicit or spontaneous breaking of a $\mathbb{Z}_2$ symmetry in the $\chi$ sector, such that (i) $A^2(\chi)$ contains a term linear in $\chi$ or (ii) we can expand $\chi=v_{\chi}+\delta\chi$ around a non-zero vev $v_{\chi}$, giving $A^2(\chi)=A^2(v_{\chi})\:+\:2A(v_{\chi})A'(v_{\chi})\delta\chi$. The Brans-Dicke theory~\cite{Brans:1961sx} and chameleon model~\cite{Khoury:2003aq,Khoury:2003rn,Khoury:2013yya} fall into the former case and the symmetron model~\cite{Hinterbichler:2010es,Hinterbichler:2011ca} (for earlier variants, see Refs.~\cite{Gessner:1992flm,Dehnen:1992rr,Damour:1994zq,Pietroni:2005pv,Olive:2007aj}) into the latter.

In order to study all of the low-dimension operators involving $\chi$ and the would-be SM Higgs field generated by the Weyl transformation explicitly, we write the coupling function in the general form
\begin{equation}
\label{eq:A}
A^2(\chi)\ =\ a\:+\:b\,\frac{\chi}{M}\:+\:c\,\frac{\chi^2}{M^2}\:+\:\mathcal{O}\bigg(\frac{\chi^3}{M^3}\bigg)\;,
\end{equation}
where $a$, $b$ and $c$ are dimensionless constants and $M$ is an energy scale. We also include a potential for the $\chi$ field
\begin{equation}
\label{eq:chipot}
V(\chi)\ =\ \frac{d}{2}\,\mu_{\chi}^2\,\chi^2\: +\:\frac{\lambda_{\chi}}{4 !}\,\chi^4\;,
\end{equation}
where $\mu_{\chi}$ is a mass, $d = \pm 1$, so that we can choose whether or not the mass term is tachyonic (allowing for spontaneous symmetry breaking in the $\chi$ sector), and $\lambda_{\chi}$ is another dimensionless constant. The non-gravitational part of the Einstein-frame Lagrangian can then be written
\begin{align}
\tilde{\mathcal{L}}\ &=\ -\:\frac{1}{2}\,\tilde{g}^{\mu\nu}\,\bigg(1\:+\:\frac{b^2\tilde{\phi}^2}{4M^2}\bigg)\,\partial_{\mu}\chi\,\partial_{\nu}\chi\:-\:\frac{1}{2}\,\tilde{g}^{\mu\nu}\,\partial_{\mu}\tilde{\phi}\,\partial_{\nu}\tilde{\phi}\nonumber\\&\qquad+\:\frac{1}{2}\,\tilde{g}^{\mu\nu}\,\left(b+2ac\,\frac{\chi}{M}-b^2\,\frac{\chi}{M}\right)\frac{\tilde{\phi}}{M}\,\partial_{\mu}\tilde{\phi}\,\partial_{\nu}\chi\:+\:\frac{1}{2}\,\mu^2\,\tilde{\phi}^2\left(a+b\,\frac{\chi}{M}+c\,\frac{\chi^2}{M^2}\right)\nonumber\\&\qquad-\:\frac{\lambda}{4!}\,\tilde{\phi}^4\:-\:\frac{3}{2}\,\frac{\mu^4}{\lambda}\left(a+2ab\,\frac{\chi}{M}+2ac\,\frac{\chi^2}{M^2}+b^2\,\frac{\chi^2}{M^2}\right)\nonumber\\&\qquad-\:\frac{d}{2}\,\mu_{\chi}^2\chi^2\:-\:\frac{\lambda_{\chi}}{4!}\,\chi^4\:-\:\bar{\tilde{\psi}}i\overset{\leftrightarrow}{\tilde{\slashed{\partial}}}\tilde{\psi}\:-\:y\,\bar{\tilde{\psi}}\tilde{\phi}\tilde{\psi}\:+\:\mathcal{O}\big(\chi^3/M^3\big)\;.
\end{align}
Defining
\begin{equation}
\label{eq:chiredef}
\tilde{\chi}\ \equiv\ \bigg(1\:+\:\frac{b^2\tilde{\phi}^2}{4M^2}\bigg)^{1/2}\chi\;,
\end{equation}
to approach canonical normalization for the $\chi$ field, we have (keeping terms up to order $\tilde{\chi}^2/M^2$ and $\tilde{\phi^2}/M^2$)
\begin{align}
\label{eq:EFTL}
\tilde{\mathcal{L}}\ &=\ -\:\frac{1}{2}\,\tilde{g}^{\mu\nu}\,\partial_{\mu}\tilde{\chi}\,\partial_{\nu}\tilde{\chi}\:-\:\frac{1}{2}\,\tilde{g}^{\mu\nu}\,\partial_{\mu}\tilde{\phi}\,\partial_{\nu}\tilde{\phi}\:+\:\frac{1}{2}\,\tilde{g}^{\mu\nu}\,\frac{\tilde{\phi}}{M}\left(b+2ac\,\frac{\tilde{\chi}}{M}-b^2\,\frac{\tilde{\chi}}{2 M}\right)\,\partial_{\mu}\tilde{\phi}\,\partial_{\nu}\tilde{\chi}\nonumber\\&\qquad+\:\frac{1}{2}\,\mu^2\,\tilde{\phi}^2\left(a+b\,\frac{\tilde{\chi}}{M}+c\,\frac{\tilde{\chi}^2}{M^2}\right)\:-\:\frac{\lambda}{4!}\,\tilde{\phi}^4\:-\:\frac{3}{2}\,\frac{\mu^4}{\lambda}\left(a+2ab\,\frac{\tilde{\chi}}{M}+2ac\,\frac{\tilde{\chi}^2}{M^2}+b^2\,\frac{\tilde{\chi}^2}{M^2}\right)\nonumber\\&\qquad-\:\frac{d}{2}\mu_{\chi}^2\tilde{\chi}^2\left(1-\frac{b^2\tilde{\phi}^2}{4 M^2}\right)\:-\:\frac{\lambda_{\chi}}{4!}\,\tilde{\chi}^4\left(1-\frac{b^2\tilde{\phi}^2}{2M^2}\right)\:-\:\bar{\tilde{\psi}}i\overset{\leftrightarrow}{\tilde{\slashed{\partial}}}\tilde{\psi}\:-\:y\,\bar{\tilde{\psi}}\tilde{\phi}\tilde{\psi}\:+\:\cdots\;.
\end{align}
The resulting Einstein-frame theory is nothing other than a Higgs-portal theory. In this sense, there can be little distinction between modifications of general relativity involving conformally coupled scalars and scalar extensions of the SM. Specifically, the only way to couple additional singlet scalar fields into the SM is via precisely the operators that can be generated by the Weyl transformation of a conformally coupled theory.

The Lagrangian in Eq.~\eqref{eq:EFTL} is that of an effective field theory with a cut-off scale given by $M$, and we have kept terms up to second order in $\{\phi,\chi\}/M$, assuming that all other mass scales are much smaller than $M$. This has left us with a combination of dimension-four, -five and -six operators. However, in order to understand the origin of any fifth forces, mediated by the conformally coupled scalar $\chi$, it is sufficient for us to consider only the dimension-four operators generated by the Weyl transformation in the low-energy, symmetry-broken theory. In other words, throughout the remainder of this work, we study low-energy theories whose dimension-four operators are fixed by requiring that they originate from the Weyl transformation of a particular conformally coupled theory, that is the couplings are fixed at a given energy scale to be those arising from the Weyl transformation of a conformally coupled scalar-tensor theory. We remark that it is, in fact, only at a fixed scale that we can define the Einstein frame, wherein all fields are minimally coupled to gravity. Following the renormalization-group evolution to any other scale, the non-minimal couplings will be regenerated, as occurs for the SM Higgs (see, e.g., Ref.~\cite{Herranen:2014cua}).

In the following subsections, we will treat two prototypal conformally coupled models in detail. In particular, we will be careful to clarify in each case how the scalar modes mix, and which mode(s) couple to the matter fermions directly.  When appropriate to do so, we hereafter suppress terms involving the fermion fields and neglect corrections to the self-interactions of the conformally coupled scalar $\chi$, since the latter amounts only to a redefinition of the couplings.  All of the derived expressions are correct to lowest order in $\{\mu,v_{\phi}\equiv\braket{\phi},v_{\chi}\equiv\braket{\chi}\}/M$.


\subsection{Brans-Dicke (Chameleon) theory}
\label{sec:cham1}

We first consider the simplest and most well-studied scalar-tensor theory. This is  commonly written in the form of a Brans-Dicke theory~\cite{Brans:1961sx}, whose Jordan-frame action is
\begin{equation}
\label{eq:BDJF}
S\ =\ \int\!{\rm d}^4x\;\sqrt{-\,g}\bigg[\frac{X}{2}\,\mathcal{R}\:-\:\frac{\omega(X)}{2X}\,g^{\mu\nu}\,\partial_{\mu}X\,\partial_{\nu}X\bigg]\:+\:S_{\rm SM}[g_{\mu\nu},\{\psi\}]\;.
\end{equation}
Note that $X$ has mass dimension 2 and, for aesthetic reasons, we have used a non-standard normalization for the Brans-Dicke scalar $X$. Alternatively, we can transform the action to the Einstein frame for calculational simplicity, wherein it becomes
\begin{align}
S\ =\ \int\!{\rm d}^4x\;\sqrt{-\,\tilde{g}}\bigg[\frac{M_{\rm Pl}^2}{2}\,\tilde{\mathcal{R}}\:-\:\frac{2\omega(X)+3}{4X}\,\frac{M_{\rm Pl}^2}{X}\,\tilde{g}^{\mu\nu}\,\partial_{\mu}X\,\partial_{\nu}X\bigg]\:+\:S_{\rm SM}[A^2(\chi)\tilde{g}_{\mu\nu},\{\psi\}]\;,
\end{align}
where $\chi$ is the canonically-normalized field (neglecting terms of order $\tilde{\phi}^2/M_{\rm Pl}^2$ that arise from the kinetic mixing between $X$ and $\phi$):
\begin{equation}
\chi\ \equiv\ M_{\rm Pl}\int^{M_{\rm Pl}^2}_{X}\frac{{\rm d}X'}{X'}\;\sqrt{\frac{2\omega(X')+3}{2}}\;.
\end{equation}
In order to proceed analytically, we assume $\omega(X)={\rm const.}$, such that
\begin{equation}
\chi\ =\ -\:\sqrt{\frac{2\omega+3}{2}}\,M_{\rm Pl}\,\ln\frac{X}{M_{\rm Pl}^2}\;,\qquad X\ =\ M_{\rm Pl}^2\,\exp\bigg[-\:\sqrt{\frac{2}{2\omega+3}}\,\frac{\chi}{M_{\rm Pl}}\bigg]\;.
\end{equation}
Given the form of the Jordan-frame non-minimal coupling $\mathcal{L}\supset X\mathcal{R}/2$ in Eq.~\eqref{eq:BDJF}, we therefore have
\begin{equation}
\label{eq:BDcouplingfunc}
A^2(\chi)\ =\ \frac{M_{\rm Pl}^2}{X}\ =\ \exp\bigg[2\,\frac{\chi}{M}\bigg]\;,
\end{equation}
where
\begin{equation}
\label{eq:M2Mpl}
M^2\ \equiv\ 2(2\omega+3)M_{\rm Pl}^2\;.
\end{equation}
Thus, this Brans-Dicke theory is equivalent to taking $a=1$, $b=2$, $c=2$ in Eq.~\eqref{eq:A} and setting the $\chi$ potential to zero, i.e.~taking $d=0$ and $\lambda_{\chi}=0$ in Eq.~\eqref{eq:chipot}.  From Eq.~\eqref{eq:EFTL}, we then have
\begin{align}
\label{eq:BDL}
\tilde{\mathcal{L}}\ &=\ -\:\frac{1}{2}\,\tilde{g}^{\mu\nu}\,\partial_{\mu}\tilde{\chi}\,\partial_{\nu}\tilde{\chi}\:-\:\frac{1}{2}\,\tilde{g}^{\mu\nu}\,\partial_{\mu}\tilde{\phi}\,\partial_{\nu}\tilde{\phi}\:+\:\tilde{g}^{\mu\nu}\,\frac{\tilde{\phi}}{M}\left(1+\frac{\tilde{\chi}}{M}\right)\,\partial_{\mu}\tilde{\phi}\,\partial_{\nu}\tilde{\chi}\nonumber\\&\qquad+\:\frac{1}{2}\,\mu^2\,\tilde{\phi}^2\left(1+2\,\frac{\tilde{\chi}}{M}+2\,\frac{\tilde{\chi}^2}{M^2}\right)\:-\:\frac{\lambda}{4!}\,\tilde{\phi}^4\:-\:\frac{3}{2}\,\frac{\mu^4}{\lambda}\left(1+4\,\frac{\tilde{\chi}}{M}+8\,\frac{\tilde{\chi}^2}{M^2}\right)\:+\:\dots\;.
\end{align}
Note that, in Eq.~\eqref{eq:BDL}, we have neglected the cubic and quartic self-interactions $\tilde{\mathcal{L}}\supset -16\mu^4\tilde{\chi}^3/(\lambda M^3)$ and $\tilde{\mathcal{L}}\supset -16\mu^4\tilde{\chi}^4/(\lambda M^4)$, generated by the Jordan-frame cosmological constant. If we were to account for these terms, we would, in fact, arrive at a variant of the quartic chameleon model~\cite{Gubser:2004uf}, and we will therefore refer to the present model as a chameleon theory. The generation of self-interactions for the conformally coupled field means that these models will generically possess screening mechanisms that could dynamically suppress the force in regions of high density, unless the original Jordan-frame couplings are fine-tuned to remove them. The details of any screening are not the focus of this work, and we leave the study of viable screened Higgs-portal models for future work~\cite{inprep}.

In order to understand the origins and behaviour of any fifth forces, we now turn our attention to the low-energy, symmetry-broken theory. The global minima of the scalar potential lie at
\begin{gather}
v_{\phi}\ =\ \pm\,\frac{\sqrt{6}\mu}{\sqrt{\lambda}}\;\qquad v_{\chi}\ =\ 0\;.
\end{gather}
The Lagrangian describing the low-energy degrees of freedom can then be found straightforwardly by shifting $\tilde{\phi}\to v_{\phi}+\tilde{\phi}$ in Eq.~\eqref{eq:BDL}, and we obtain
\begin{align}
\label{eq:chamlag2}
\tilde{\mathcal{L}}\ &=\ -\:\frac{1}{2}\,\tilde{g}^{\mu\nu}\,\partial_{\mu}\tilde{\chi}\,\partial_{\nu}\tilde{\chi}\:-\:\frac{1}{2}\,\tilde{g}^{\mu\nu}\,\partial_{\mu}\tilde{\phi}\,\partial_{\nu}\tilde{\phi}\:+\:\tilde{g}^{\mu\nu}\,\frac{v_{\phi}}{M}\,\partial_{\mu}\tilde{\phi}\,\partial_{\nu}\tilde{\chi}\nonumber\\&\qquad-\:\frac{1}{2}\,2\,\mu^2\,\tilde{\phi}^2\:+\:2\mu^2\,\frac{v_{\phi}}{M}\,\tilde{\phi}\,\tilde{\chi}\:-\:\frac{1}{2}\,2\mu^2\,\frac{v_{\phi}^2}{M^2}\,\tilde{\chi}^2\nonumber\\&\qquad-\:\frac{\lambda}{3!}\,v_{\phi}\,\tilde{\phi}^3\:-\:\frac{\lambda}{4!}\,\tilde{\phi}^4\nonumber\\&\qquad+\:\frac{\mu^2}{M}\,\tilde{\phi}^2\tilde{\chi}\:+\:\frac{\mu^2}{M^2}\tilde{\phi}^2\tilde{\chi}^2\:+\:2\frac{\mu^2}{M^2}\,v_{\phi}\tilde{\phi}\tilde{\chi}^2\:+\:\dots\;,
\end{align}
where the ellipsis also includes the self-interactions of the $\tilde{\chi}$ field. In terms of the Higgs-chameleon interactions, we have been left with a kinetic mixing term (line one), a mass mixing (line two) and Higgs-portal terms (line four).

As we will see in Sec.~\ref{sec:fifthforce}, the mass mixing leads to the dominant long-range fifth force. The squared mass matrix has the form
\begin{equation}
\mathbf{m}^2\ =\ 2\mu^2\begin{pmatrix} 1 & -\:\frac{v_{\phi}}{M} \\ -\:\frac{v_{\phi}}{M} & \frac{v_{\phi}^2}{M^2}\end{pmatrix}\;,
\end{equation}
with eigenvalues
\begin{equation}
m^2_h\ =\ 2\mu^2\bigg(1\:+\:\frac{v^2_{\phi}}{M^2}\bigg)\;,\qquad m^2_{\zeta}\ =\ 0\;.
\end{equation}
As one would expect, we have a massive mode $h$ (the Higgs\footnote{Extending this to the SM, the corresponding analysis could be made straightforwardly in unitary gauge after electroweak symmetry breaking.}) and a massless mode $\zeta$ (the chameleon). However, because of this mass mixing, we have
\begin{equation}
\tilde{\phi}\ =\ h\:+\:\frac{v_{\phi}}{M}\,\zeta\;,
\end{equation}
such that both the heavy and light modes couple to the SM fermions via the Yukawa interaction in Eq.~\eqref{eq:EFTL}. In particular, the light chameleon mode couples as
\begin{equation}
\label{eq:chamcoup}
\mathcal{L}\ \supset\ -\:\frac{\zeta}{M}\,m\bar{\tilde{\psi}}\tilde{\psi}\ = \ \frac{\zeta}{M}\,T_{\psi}^{\rm OS}\;,
\end{equation}
where $m=yv_{\phi}$. This is precisely the standard chameleon coupling to the trace of the on-shell energy-momentum tensor of a fermion with Dirac mass $m$. Notice that it is \emph{not} the original non-minimally coupled field $\chi$ that couples to the fermion energy-momentum tensor, as in the standard arguments presented at the beginning of this section [cf.~Eq.~\eqref{eq:usualfifth}], but rather the light mode $\zeta=\tilde{\chi}+(v_{\phi}/M)\tilde{\phi}$.


\subsection{Symmetron}
\label{sec:sym1}

Going one step beyond the minimal scalar-tensor theory discussed in the preceding subsection, we turn our attention to the symmetron model, wherein the conformally coupled sector also exhibits spontaneous symmetry breaking. The symmetron corresponds to choosing $a=1$, $b = 0$, $c=1$ and $d=-1$ in Eqs.~\eqref{eq:A}, \eqref{eq:chipot} and~\eqref{eq:EFTL}, giving the Einstein-frame Lagrangian
\begin{align}
\label{eq:symlag1}
\tilde{\mathcal{L}}\ &=\ -\:\frac{1}{2}\,\tilde{g}^{\mu\nu}\,\partial_{\mu}\tilde{\chi}\,\partial_{\nu}\tilde{\chi}\:-\:\frac{1}{2}\,\tilde{g}^{\mu\nu}\,\partial_{\mu}\tilde{\phi}\,\partial_{\nu}\tilde{\phi}\:+\:\tilde{g}^{\mu\nu}\,\frac{\tilde{\phi}\tilde{\chi}}{M^2}\,\partial_{\mu}\tilde{\phi}\,\partial_{\nu}\tilde{\chi}\nonumber\\&\qquad+\:\frac{1}{2}\,\mu^2\,\tilde{\phi}^2\left(1+\frac{\tilde{\chi}^2}{M^2}\right)\:-\:\frac{\lambda}{4!}\,\tilde{\phi}^4\:-\:\frac{3}{2}\,\frac{\mu^4}{\lambda}\left(1+2\,\frac{\tilde{\chi}^2}{M^2}\right)\nonumber\\&\qquad+\:\frac{1}{2}\,\mu_{\chi}^2\tilde{\chi}^2\:-\:\frac{\lambda_{\chi}}{4!}\,\tilde{\chi}^4\:+\:\dots\;.
\end{align}
In this case, we have neglected corrections to the quartic self-interaction of order $\mu^4/M^4$, which are again generated by the Jordan-frame cosmological constant.

For this symmetron model, the global minima of the potential lie at
\begin{gather}
v_{\phi}\ =\ \pm\,\frac{\sqrt{6}\mu}{\sqrt{\lambda}}\bigg(1+\frac{v_{\chi}^2}{2M^2}\bigg)\;\qquad v_{\chi}\ =\ \pm'\frac{\sqrt{6}\mu_{\chi}}{\sqrt{\lambda_{\chi}}}\;.
\end{gather}
(The prime on the second $\pm$ indicates that the sign of $v_{\chi}$ is independent to that of $v_{\phi}$, i.e.~there are four degenerate minima.) As in Subsec.~\ref{sec:cham1}, we shift $\tilde{\phi}\to v_{\phi}+\tilde{\phi}$ and $\chi\to v_{\chi}+\tilde{\chi}$ in Eq.~\eqref{eq:symlag1}, and the Lagrangian for the fluctuations can be written
\begin{align}
\label{eq:symlag2}
\tilde{\mathcal{L}}\ &=\ -\:\frac{1}{2}\,\tilde{g}^{\mu\nu}\,\partial_{\mu}\tilde{\chi}\,\partial_{\nu}\tilde{\chi}\:-\:\frac{1}{2}\,\tilde{g}^{\mu\nu}\,\partial_{\mu}\tilde{\phi}\,\partial_{\nu}\tilde{\phi}\:+\:\tilde{g}^{\mu\nu}\,\frac{v_{\phi}v_{\chi}}{M^2}\,\partial_{\mu}\tilde{\phi}\,\partial_{\nu}\tilde{\chi}\nonumber\\&\qquad-\:\frac{1}{2}\,2\,\mu^2\bigg(1\:+\:\frac{v_{\chi}^2}{M^2}\bigg)\,\tilde{\phi}^2\:+\:2\mu^2\,\frac{v_{\phi}v_{\chi}}{M^2}\,\tilde{\phi}\,\tilde{\chi}\:-\:\frac{1}{2}\,2\mu_{\chi}^2\,\tilde{\chi}^2\nonumber\\&\qquad-\:\frac{\lambda}{3!}\,v_{\phi}\,\tilde{\phi}^3\:-\:\frac{\lambda}{4!}\,\tilde{\phi}^4\:-\:\frac{\lambda_{\chi}}{3!}\,v_{\chi}\,\tilde{\chi}^3\:-\:\frac{\lambda_{\chi}}{4!}\,\tilde{\chi}^4\nonumber\\&\qquad+\:\frac{\mu^2}{M^2}\,v_{\chi}\,\tilde{\phi}^2\tilde{\chi}\:+\:\frac{1}{2}\,\frac{\mu^2}{M^2}\,\tilde{\phi}^2\tilde{\chi}^2\:+\:\frac{\mu^2}{M^2}\,v_{\phi}\tilde{\phi}\tilde{\chi}^2\:+\:\dots\;.
\end{align}
Much like the chameleon case, the Higgs-symmetron interactions comprise a kinetic mixing term (line one), a mass mixing (line two) and Higgs-portal terms (line four). The squared mass matrix has the form
\begin{equation}
\mathbf{m}^2\ =\ \begin{pmatrix} 2\mu^2\Big(1+\frac{v_{\chi}^2}{M^2}\Big)  & -\:2\mu^2\,\frac{v_{\chi}v_{\phi}}{M^2} \\  -\:2\mu^2\,\frac{v_{\chi}v_{\phi}}{M^2} & 2\mu_{\chi}^2\end{pmatrix}\;,
\end{equation}
with eigenvalues
\begin{equation}
m^2_h\ =\ 2\mu^2\bigg(1\:+\:\frac{v_{\chi}^2}{M^2}\bigg)\;,\qquad m^2_{\zeta}\ =\ 2\mu_{\chi}^2\;.
\end{equation}
As in the chameleon case, we find  a massive mode $h$ (the Higgs) and a light mode $\zeta$ (the symmetron), with
\begin{equation}
\tilde{\phi}\ =\ h\:+\:\frac{v_{\phi}v_{\chi}}{M^2}\,\zeta\;.
\end{equation}
The light mode therefore couples to the fermion mass term as
\begin{equation}
\label{eq:symcoup}
\mathcal{L}\ \supset\ -\:\frac{v_{\chi}\zeta}{M^2}\,m\bar{\tilde{\psi}}\tilde{\psi}\ = \ \frac{v_{\chi}\zeta}{M^2}\,T_{\psi}^{\rm OS}\;,
\end{equation}
where $m=yv_{\phi}$, and this is again the standard symmetron coupling to the trace of the on-shell energy-momentum tensor of a fermion with Dirac mass $m$.


\section{Fifth forces}
\label{sec:fifthforce}

Having understood the interactions that arise between the scalar field $\chi$, the would-be Higgs field $\phi$ and the fermionic fields in the symmetry-broken theory, we are now able to isolate the various potential sources of fifth forces between the SM fermions. In this section, we compute the leading, tree-level fifth forces from each of these sources. Most importantly, and in order to make clear the connection with Higgs-portal theories and to emphasize the importance of explicit scale-breaking terms, we treat the contributions from the kinetic and mass mixings separately. Doing so will allow us to show that it is the mass mixing which dominates long-range fifth forces. A convenient way to determine the relevant Yukawa potential is then to consider the corrections to the non-relativistic limit of the Higgs-mediated M\o ller scattering ($e^-e^-\to e^-e^-$) from the conformally coupled scalar.
One could, of course, proceed alternatively by perturbatively diagonalizing the mass and kinetic terms; however, doing so prevents a comparison of the relative contributions from the mass and kinetic mixings. We need only consider the $t$-channel exchange and assume the scattering electrons to be distinguishable. We will consider each type of interaction in turn.


\subsection{Higgs portal}

The Higgs-portal terms in Eqs.~\eqref{eq:chamlag2} and~\eqref{eq:symlag2} have the generic form
\begin{equation}
\tilde{\mathcal{L}}_P\ =\ \frac{1}{2}\,\alpha_{P12}\,\tilde{\phi}\,\tilde{\chi}^2\:+\: \frac{1}{4}\,\alpha_{P22}\,\tilde{\phi}^2\,\tilde{\chi}^2\:+\:\frac{1}{2}\,\alpha_{P21}\,\tilde{\phi}^2\,\tilde{\chi}\;.
\end{equation}
Since the Yukawa coupling of the Higgs field to the fermions is linear in the Higgs field, these couplings lead only to loop corrections to the scattering of the fermions, which we therefore neglect relative to the tree-level scatterings that follow.


\subsection{Mass mixing}
\label{sec:massmixing}

The mass mixing is generically of the form
\begin{equation}
\tilde{\mathcal{L}}_M\ =\ \alpha_M\,\tilde{\phi}\,\tilde{\chi}\;.
\end{equation}
The corresponding  matrix element describing the correction to  the M\o ller scattering  is
\begin{align}
i\mathcal{M}(e^-e^-\to e^-e^-)\: &\supset\: \bar{u}(\mathbf{p}_1,s_1)(-\,iy)u(\mathbf{p}_3,s_3)\nonumber\\&\qquad\times\:\frac{i}{t-m_{\phi}^2}\Bigg[\sum_{n\,=\,0}^{\infty}(i\alpha_M)^{2n}\bigg(\frac{i}{t-m_{\phi}^{2}}\bigg)^{\!n}\bigg(\frac{i}{t-m_{\chi}^2}\bigg)^{\!n}\Bigg]
\nonumber\\&\qquad\times\:
\bar{u}(\mathbf{p}_2,s_2)(-\,iy)u(\mathbf{p}_4,s_4)\;,
\end{align}
where $t=-\,(p_1-p_3)^2$ is the usual Mandelstam variable,\footnote{Recall that $p_1^2=-\,m_{e}^2$ on-shell for our signature conventions.} $u(\mathbf{p},s)$ is a four-spinor of spin projection $s$ and $\bar{u}(\mathbf{p},s)$ is its Dirac conjugate. 
In addition, $m_{\phi}^2$ and $m_{\chi}^2$ are the second variations of the action with respect to \smash{$\tilde{\phi}$} and \smash{$\tilde{\chi}$} in the symmetry-broken phase(s). Performing the summation over $n$ and using the fact that $\bar{u}(\mathbf{p},s)u(\mathbf{q},s')=2m_e\delta_{ss'}$, where $m_e$ is the electron mass, we find
\begin{align}
i\mathcal{M}(e^-e^-\to e^-e^-)\ &\supset\ -4iy^2m_e^2\delta_{s_1s_3}\delta_{s_2s_4}\,\frac{t-m_{\chi}^2}{(t-m_{\chi}^2)(t-m_{\phi}^2)-\alpha_M^2}\;.
\end{align}
In the non-relativistic approximation, we take $t=-\,\mathbf{Q}^2$, and the Yukawa potential has the form
\begin{align}
V(r)\ &=\  -\:y^2\int\!\frac{{\rm d}^3\mathbf{Q}}{(2\pi)^3}\;e^{i\mathbf{Q}\cdot \mathbf{x}}\frac{\mathbf{Q}^2+m_{\chi}^2}{(\mathbf{Q}^2+m_{\chi}^2)(\mathbf{Q}^2+m_{\phi}^2)-\alpha_M^2}\nonumber\\ &\approx\ -\:\frac{y^2}{4\pi}\,\bigg(1\:-\:\frac{\alpha_M^2}{m_{\phi}^4}\bigg)\,\frac{e^{-m_h r}}{r}\:-\:\frac{y^2}{4\pi}\,\frac{\alpha_M^2}{m_{\phi}^4}\,\frac{e^{-m_{\zeta}r}}{r}\;,
\label{eq:brokenYukawa}
\end{align}
where we have expanded the coefficients of each contribution to the potential to leading order in $\alpha_M^2$. The fifth force due to the light mode $\zeta$ is now present. Whilst this would appear to be suppressed by four powers of the mass $m_{\phi}\sim \sqrt{2}\mu$, this is in fact not the case, since $\alpha_M\propto \mu^2$. For the chameleon and symmetron cases, we therefore find
\begin{equation}
\label{eq:fifthforce}
V(r)\ \supset\ -\:\frac{1}{4\pi}\,\frac{m_e^2}{M^2}\,\frac{1}{r}\begin{cases} 1\;,&\qquad \text{chameleon}\;,\\
\displaystyle{\frac{v_{\chi}^2}{M^2}}\,e^{-m_{\zeta}r}\;,&\qquad \text{symmetron}\;,\end{cases}
\end{equation}
recovering the fifth force consistent with Eqs.~\eqref{eq:chamcoup} and \eqref{eq:symcoup}.


\subsection{Kinetic mixing}

The kinetic mixing terms have the generic form
\begin{equation}
\tilde{\mathcal{L}}_{K}\ =\ \alpha_{K}\,\tilde{g}^{\mu\nu}\,\partial_{\mu}\tilde{\phi}\,\partial_{\nu}\tilde{\chi}\;.
\end{equation}
At leading order, the corresponding matrix element in the M\o ller scattering is
\begin{align}
i\mathcal{M}(e^-e^-\to e^-e^-)\ &\supset\ \bar{u}(\mathbf{p}_1,s_1)(-\,iy)u(\mathbf{p}_3,s_3)\,\frac{i}{t-m_{\phi}^2}(i\alpha_{K} t)\frac{i}{t-m_{\chi}^2}(i\alpha_{K} t)\frac{i}{t-m_{\phi}^2}\nonumber\\&\qquad\times\:
\bar{u}(\mathbf{p}_2,s_2)(-\,iy)u(\mathbf{p}_4,s_4)\;.
\end{align}
We see immediately that the factors of $t$ in the numerator, which arise from the derivative coupling, would cancel any massless pole from the conformally coupled scalar, cf.~the Higgs-dilaton case in Sec.~\ref{sec:HiggsDilaton}. Again using the fact that $\bar{u}(\mathbf{p},s)u(\mathbf{q},s')=2m_e\delta_{ss'}$, we find
\begin{align}
i\mathcal{M}(e^-e^-\to e^-e^-)\ &\supset\ -4iy^2m_e^2\delta_{s_1s_3}\delta_{s_2s_4}\,\frac{\alpha_{K}^2t^2}{(t-m_{\phi}^2)^2(t-m_{\chi}^2)}\;.
\end{align}
In the non-relativistic approximation, the contribution to the Yukawa potential is given by
\begin{align}
V(r)\ &=\ -\:y^2\int\!\frac{{\rm d}^3\mathbf{Q}}{(2\pi)^3}\;e^{i\mathbf{Q}\cdot \mathbf{x}}\,\frac{\alpha_{K}^2\mathbf{Q}^4}{(\mathbf{Q}^2+m_{\phi}^2)^2(\mathbf{Q}^2+m_{\chi}^2)}\nonumber\\ &\approx\ -\:\frac{y^2}{8\pi}\,\alpha_{K}^2\,\frac{e^{-\,m_{\phi}r}}{r}\,\bigg(1\:-\:\frac{m_{\phi}r}{2}\bigg)\:-\:\frac{y^2}{4\pi}\,\alpha_{K}^2\,\frac{m_{\chi}^4}{m_{\phi}^4}\,\frac{e^{-\,m_{\chi}r}}{r}\;,
\end{align}
where we have this time expanded the coefficients of each contribution to the potential to leading order in $\alpha_{K}^2$. For the chameleon and symmetron cases, we therefore find
\begin{equation}
\label{eq:Yukkin}
V(r)\ \supset\ -\:\frac{1}{4\pi}\,\frac{m_e^2}{M^2}\,\frac{m_{\chi}^4}{m_{\phi}^4}\,\frac{e^{-m_{\chi}r}}{r}\begin{cases} 1\;,&\qquad \text{chameleon}\;,\\
\displaystyle{\frac{v_{\chi}^2}{M^2}}\;,&\qquad \text{symmetron}\;.\end{cases}
\end{equation}
We see that the kinetic mixing contribution to the potential has  an additional suppression relative to the contribution from the mass mixing in Eq.~\eqref{eq:fifthforce} by a factor of $m_{\chi}^4/m_{\phi}^4$ ($m_{\chi}<m_{\phi}$), and the mass mixing therefore dominates any long-range fifth force. 


\section{Explicit scale breaking}
\label{sec:ExplicitScale}

In the previous section, we have seen that the dominant fifth force arises from the mass mixing between the would-be Higgs and the conformally coupled scalar. This mixing is present because of the mass term ($\propto \mu^2$) of the Higgs field, which provides the only source of explicit scale breaking in the SM.  We have also seen that the Higgs mass parameter cancels in the final result for the fifth force, and the implication of removing all, or part, of the Higgs mass term and instead generating it through the spontaneous breaking of scale symmetry is clear. Namely, the fifth force will be suppressed by a ratio of the explicit scale-breaking mass to the total mass of the Higgs field, vanishing when the explicit scale-breaking mass is set to zero. As we will see in Sec.~\ref{sec:HiggsDilaton}, the contribution from the kinetic mixing also vanishes in the absence of any explicit scale-breaking masses. Specifically, as we saw in Eq.~\eqref{eq:Yukkin}, the fifth force vanishes when $m_{\chi}^2\to 0$ (for a finite Higgs mass), and this is precisely what happens in scale-invariant extensions of the SM that contain a massless dilaton.

In this section, we illustrate the above suppression of the fifth force by constructing a toy realization of this situation in which only part of the Higgs mass arises from an explicit scale-breaking term. Moreover, we show how our arguments also apply to baryonic matter, in spite of the fact that the dominant contribution to the baryonic mass arises from chiral symmetry breaking. In so doing, we will find that Solar System constraints on fifth forces can, quite remarkably, be interpreted as providing an upper bound on any explicit scale breaking present in the Higgs sector of the SM.

To this end, we introduce a third scalar field $\theta$, which couples to the  Higgs field $\phi$ via the potential
\begin{equation}
U(\phi,\theta)\ =\ \frac{\lambda}{4!}\,\bigg(\phi^2-\frac{\beta}{\lambda}\,\theta^2\bigg)^2\:-\:\frac{1}{2}\,\mu^2\,\bigg(\phi^2-\frac{\beta}{\lambda}\,\theta^2\bigg)\:+\:\frac{3}{2}\,\frac{\mu^4}{\lambda}\;.
\end{equation}
By taking the Jordan-frame Lagrangian of the field $\theta$ to be
\begin{equation}
\mathcal{L}_{\theta}\ =\ -\:\frac{1}{2}\,g^{\mu\nu}\,\partial_{\mu}\theta\,\partial_{\nu}\theta\:+\:\frac{1}{2}\,\mu_{\theta}^2\,A^{-2}(\chi)\,\theta^2\:-\:\frac{\lambda_{\theta}}{4!}\,\theta^4\:-\:\frac{3}{2}\,\frac{\mu_{\theta}^4}{\lambda_{\theta}}\,A^{-4}(\chi)\;,
\end{equation}
we can ensure that no mass mixing between the $\chi$ and $\theta$ fields is generated in the Einstein frame from the additional scale-breaking parameter $\mu_{\theta}^2$. There will, of course, arise kinetic mixings, but these will play a subdominant role in the long-range fifth forces. While this particular tuning of the couplings allows us to realize a concrete scenario in which the dynamics of the $\theta$ field are stabilized and can be decoupled from the low-energy dynamics, our conclusions hold more generally [cf.~Sec.~\ref{sec:HiggsDilaton}]. After making the Weyl transformation to the Einstein frame [cf.~Eqs.~\eqref{eq:phiredef} and \eqref{eq:chiredef}], the potential for all three fields takes the form
\begin{align}
&\tilde{U}(\tilde{\phi},\tilde{\theta},\tilde{\chi})\ = \ \frac{\lambda}{4!}\,\bigg(\tilde{\phi}^2-\frac{\beta}{\lambda}\,\tilde{\theta}^2\bigg)^2\:-\:\frac{1}{2}\,\mu^2\,\bigg(\tilde{\phi}^2-\frac{\beta}{\lambda}\,\tilde{\theta}^2\bigg)\bigg(a+b\,\frac{\tilde{\chi}}{M}+c\,\frac{\tilde{\chi}^2}{M^2}\bigg)\nonumber\\&\qquad+\:\frac{3}{2}\,\frac{\mu^4}{\lambda}\bigg(a+2ab\,\frac{\tilde{\chi}}{M}+2ac\,\frac{\tilde{\chi}^2}{M^2}+b^2\,\frac{\tilde{\chi}^2}{M^2}\bigg)\:+\:\frac{d}{2}\,\mu_{\chi}^2\,\tilde{\chi}^2\bigg(1-\frac{b^2\tilde{\phi}^2}{4M^2}\bigg)\:+\:\frac{\lambda_{\chi}}{4!}\,\tilde{\chi}^4\nonumber\\&\qquad -\:\frac{1}{2}\,\mu_{\theta}^2\,\tilde{\theta}^2\:+\:\frac{\lambda_{\theta}}{4!}\,\tilde{\theta}^4\:+\:\frac{3}{2}\,\frac{\mu_{\theta}^4}{\lambda_{\theta}}\;.
\label{eq:fullpotential}
\end{align}
In the limit $\mu^2\rightarrow0$, the potential of the Higgs field becomes scale invariant, and the spontaneous symmetry breaking in the $\theta$ sector sources the required scale breaking in the Higgs sector. The (toy) SM of the preceding sections is recovered by taking the constant $\beta \to 0$. For finite $\beta$ and $\mu^2$, the scale breaking in the Higgs sector arises from both explicit and spontaneous sources, as we require.


\subsection{Chameleon}
\label{sec:chamexamp}

For the chameleon ($a=1$, $b=2$, $c=2$, $d=0$), Eq.~\eqref{eq:fullpotential} simplifies slightly to
\begin{align}
&\tilde{U}(\tilde{\phi},\tilde{\theta},\tilde{\chi})\ = \ \frac{\lambda}{4!}\,\bigg(\tilde{\phi}^2-\frac{\beta}{\lambda}\,\tilde{\theta}^2\bigg)^2\:-\:\frac{1}{2}\,\mu^2\,\bigg(\tilde{\phi}^2-\frac{\beta}{\lambda}\,\tilde{\theta}^2\bigg)\bigg(1+2\,\frac{\tilde{\chi}}{M}+2\,\frac{\tilde{\chi}^2}{M^2}\bigg)\nonumber\\&\qquad+\:\frac{3}{2}\,\frac{\mu^4}{\lambda}\bigg(1+4\,\frac{\tilde{\chi}}{M}+8\,\frac{\tilde{\chi}^2}{M^2}\bigg)\:+\:\frac{\lambda_{\chi}}{4!}\,\tilde{\chi}^4\:-\:\frac{1}{2}\,\mu_{\theta}^2\,\tilde{\theta}^2\:+\:\frac{\lambda_{\theta}}{4!}\,\tilde{\theta}^4\:+\:\frac{3}{2}\,\frac{\mu_{\theta}^4}{\lambda_{\theta}}\;.
\end{align}
The symmetry-breaking minima of this three-field model lie at
\begin{equation}
v_{\phi}\ =\ \pm\bigg(\frac{6\mu^2+\beta v_{\theta}^2}{\lambda}\bigg)^{1/2}\;,\qquad v_{\theta}\ =\ \pm'\: \bigg(\frac{6\mu_{\theta}^2}{\lambda_{\theta}}\bigg)^{1/2}\;,\qquad v_{\chi}\ =\ 0\;.
\end{equation}
By making $\mu_{\theta}^2\gg \mu^2$, we can introduce a hierarchy between the two heavy modes, such that it is sufficient for us to consider only the mixing between the would-be Higgs field $\tilde{\phi}$ (predominantly composed of the lighter of these two modes) and the conformally coupled scalar $\tilde{\chi}$ (dominating the lightest mode of the three). In the symmetry-broken phase, shifting $\tilde{\phi}\to v_{\phi}+\tilde{\phi}$ and $\chi\to v_{\chi}+\tilde{\chi}$, the mass mixing term remains of the form in Subsec.~\ref{sec:cham1}, i.e.
\begin{equation}
\tilde{\mathcal{L}}\ \supset\ 2\mu^2\,\frac{v_{\phi}}{M}\,\tilde{\phi}\,\tilde{\chi}\;,
\end{equation}
and the contribution of the chameleon to the Yukawa potential is therefore
\begin{equation}
V(r)\ \supset\ -\:\frac{y^2}{4\pi}\,\frac{\alpha_M^2}{m_{\phi}^4}\,\frac{1}{r}\;,
\end{equation}
as before, but with
\begin{equation}
m_{\phi}^2\ =\ 2\mu^2\:+\:\frac{\beta v_{\theta}^2}{3}\;.
\end{equation}
The coupling scales like
\begin{equation}
y^2\frac{\alpha_M^2}{m_{\phi}^4}\ =\ \frac{m_e^2}{M^2}\,\frac{4\mu^4}{m_{\phi}^4}\;.
\end{equation}
In the limit $\beta\to 0$, we recover the fifth force reported earlier in Eq.~\eqref{eq:fifthforce}. However, in the limit $\mu^2\to 0$ ($m_{\phi}^2\to\beta\,v_{\theta}^2/3$), i.e.~when there are no explicit scales in the Higgs potential and the scale symmetry itself is spontaneously broken, the fifth force vanishes. For finite $\beta$ and $\mu^2$, the fifth force is, as anticipated, suppressed by the ratio of the explicit scale-breaking mass to the mass of the Higgs field.


\subsection{Symmetron}
\label{sec:symexamp}

For the symmetron ($a=1$, $b=0$, $c=1$, $d=-1$), Eq.~\eqref{eq:fullpotential} becomes
\begin{align}
&\tilde{U}(\tilde{\phi},\tilde{\theta},\tilde{\chi})\ = \ \frac{\lambda}{4!}\,\bigg(\tilde{\phi}^2-\frac{\beta}{\lambda}\,\tilde{\theta}^2\bigg)^2\:-\:\frac{1}{2}\,\mu^2\,\bigg(\tilde{\phi}^2-\frac{\beta}{\lambda}\,\tilde{\theta}^2\bigg)\bigg(1+\frac{\tilde{\chi}^2}{M^2}\bigg)\nonumber\\&\qquad+\:\frac{3}{2}\,\frac{\mu^4}{\lambda}\bigg(1+2\,\frac{\tilde{\chi}^2}{M^2}\bigg)\:-\:\frac{1}{2}\,\mu_{\chi}^2\,\tilde{\chi}^2\:+\:\frac{\lambda_{\chi}}{4!}\,\tilde{\chi}^4-\frac{1}{2}\,\mu_{\theta}^2\,\tilde{\theta}^2+\frac{\lambda_{\theta}}{4!}\,\tilde{\theta}^4+\frac{3}{2}\,\frac{\mu_{\theta}^4}{\lambda_{\theta}}\;.
\end{align}
The symmetry-breaking minima lie (to leading order) at
\begin{equation}
v_{\phi}\ =\ \pm\bigg(\frac{6\,\mu^2+\beta\,v_{\theta}^2}{\lambda}\bigg)^{1/2}\;,\qquad v_{\theta}\ =\ \pm'\: \bigg(\frac{6\mu_{\theta}^2}{\lambda_{\theta}}\bigg)^{1/2}\;,\qquad v_{\chi}\ =\ \pm''\:\bigg(\frac{6\mu_{\chi}^2}{\lambda_{\chi}}\bigg)^{1/2}\;.
\end{equation}
Treating only the two lowest lying modes as before, and shifting $\tilde{\phi}\to v_{\phi}+\tilde{\phi}$ and $\chi\to v_{\chi}+\tilde{\chi}$ in the symmetry-broken phase, the mass mixing term again remains of the form
\begin{equation}
\tilde{\mathcal{L}}\ \supset\ 2\mu^2\,\frac{v_{\phi}v_{\chi}}{M^2}\,\tilde{\phi}\,\tilde{\chi}\;,
\end{equation}
 and, in this case, the coupling scales like
\begin{equation}
y^2\frac{\alpha_M^2}{m_{\phi}^4}\ =\ \frac{m_e^2}{M^2}\,\frac{v_{\chi}^2}{M^2}\,\frac{4\mu^4}{m_{\phi}^4}\;.
\end{equation}
As in the preceding subsection, the fifth force is suppressed when the spontaneous scale breaking dominates and vanishes entirely in the limit $\mu^2\to 0$ ($m_{\phi}^2\to\beta\,v_{\theta}^2/3$).


\subsection{Hadronic matter}
\label{sec:hadron}

Our arguments on the origin of fifth forces so far hold only for the couplings to the mass terms of the elementary SM fermions, but the majority of the baryonic mass density in the universe is, of course, due to chiral symmetry breaking in QCD. One might therefore expect that the Higgs sector has little bearing on the fifth force between baryons. This intuition is, however, incorrect.

\begin{figure}
\centering
\subfloat[][]{\includegraphics[scale=0.65]{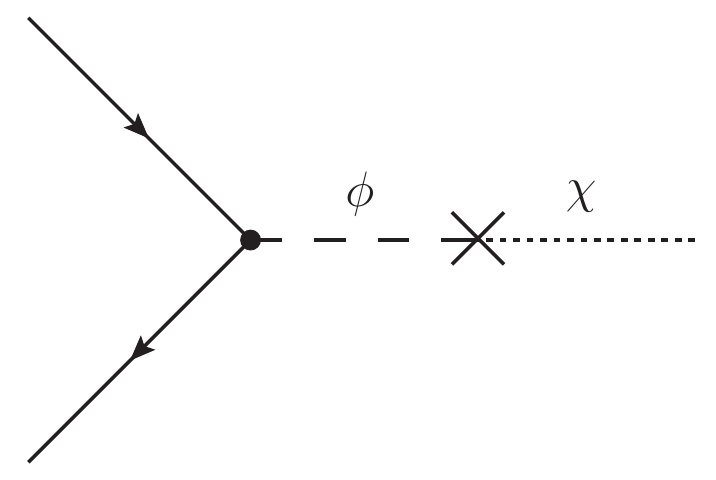}}
\subfloat[][]{\includegraphics[scale=0.65]{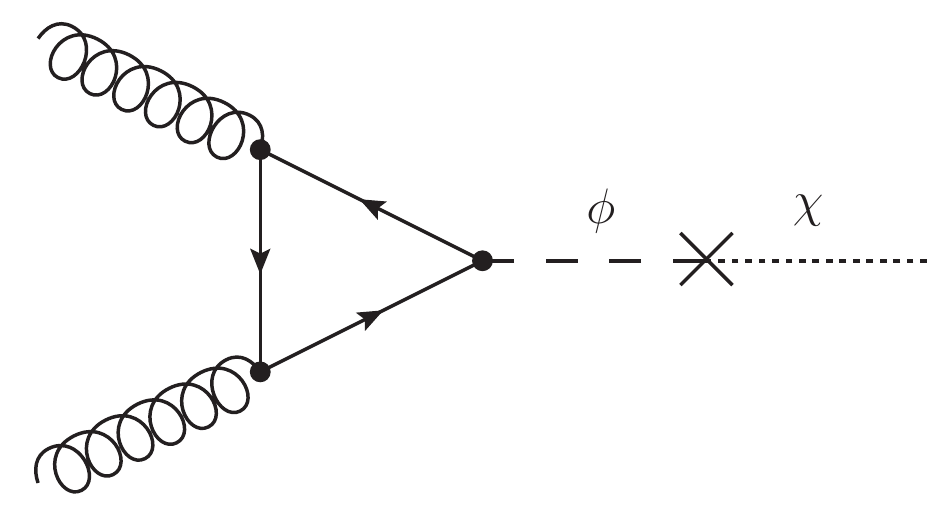}}
\subfloat[][]{\includegraphics[scale=0.65]{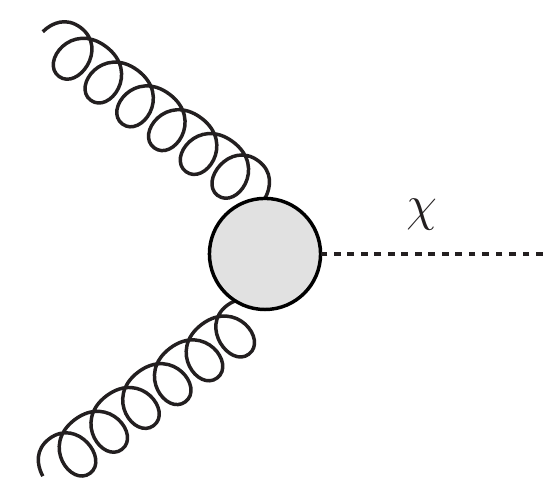}}
\caption{\label{fig:feyns} The feynman diagrams relevant to the coupling of the conformally coupled scalar to hadronic matter: (a) the coupling to fermions via their Yukawa coupling with the Higgs, (b) the coupling to gluons via the conformal anomaly and (c) the effective vertex generated by the latter. Solid lines correspond to quarks, dashed lines to the would-be Higgs field $\phi$, dotted lines to the conformally coupled scalar $\chi$ and sprung lines to gluons. The cross indicates an insertion of the mass (or kinetic) mixing.}
\end{figure}

We have seen that there are only two dimension-four operators, arising from the Weyl transformation, that couple the conformal scalar to the SM: the kinetic and mass terms of the Higgs field [cf.~Eq.~\eqref{eq:EFTL}]. We have also seen that the latter is dominant in producing the fifth force for scalars lighter than the Higgs. At dimension five, we can couple the conformal scalar to the square of the gauge field-strength tensors as
\begin{equation}
\mathcal{L}_{\rm eff}\ \supset\ -\:\frac{C}{4M}\tilde{\chi}G^a_{\mu\nu}G^{\mu\nu,a}\;,
\end{equation}
giving rise to couplings to the photons and gluons of the SM. In coupling to gluons, one might expect the scale breaking of QCD, originating via dimensional transmutation, to have a significant impact, dominating over the couplings mediated by the Higgs. However, we must fix the Wilson coefficient $C$ of this effective field theory operator by matching to the original theory. The operator of interest originates from the conformal anomaly and is therefore mediated by a fermion triangle (see Fig.~\ref{fig:feyns}). When we turn off the mixing between the conformally coupled scalar and the Higgs, the conformal scalar no longer couples to the fermion triangle, and the Wilson coefficient of this operator must therefore vanish. Without the Higgs mass term, the conformally coupled scalar is inert as far as the SM is concerned and must decouple. Hence, while chiral symmetry breaking dominates the mass of baryons, the existence of fifth forces between them nevertheless hinges on the structure of the Higgs sector.

In order to generalise our analysis to the fifth force between baryons, we first need to understand the form of the Higgs-nucleon coupling (see, e.g., Refs.~\cite{Shifman:1978zn,Gunion:1989we,Jungman:1995df}, which we follow closely). The coupling of the conformally coupled scalar to gluons via its interactions with the Higgs arises along the same lines as the coupling of neutralinos to gluons (via the same) in supersymmetric extensions of the SM~\cite{Jungman:1995df}, and the corresponding calculations can be adapted to the present context straightforwardly.

In unitary gauge, the coupling of the would-be SM Higgs field $\phi^0$ to gluons via the conformal anomaly takes the form
\begin{equation}
\mathcal{L}_{\rm eff}\ \supset\ \frac{\sqrt{2}\alpha_s}{12\pi v_{\phi}}\,N_H\,G_{\mu\nu}^aG^{\mu\nu a}\,\phi^0\;,
\end{equation}
where $G_{\mu\nu}^a$ is the gluon field strength tensor and the factor of $N_H$ counts the number of heavy quarks running in the triangle loop [see Fig.~\ref{fig:feyns}(b)]. (In the case of the Higgs-nucleon coupling, $N_H$ is commonly taken to be $3$ or $4$, depending upon whether we take the strange quark to be heavy or not, as discussed later in this section.) In the heavy quark expansion, this effective operator can be obtained from the QCD Lagrangian by making the replacement
\begin{equation}
m_{q}\bar{q}q\ \to\ -\:\frac{\alpha_s}{12\pi}\,G^{a}_{\mu\nu}G^{\mu\nu a}
\end{equation}
for each heavy quark, and the trace of the QCD energy-momentum tensor is given by
\begin{align}
\big[T^{\rm QCD}\big]_{\mu}^{\phantom{\mu}\mu}\ =\ +\:B\,\frac{\alpha_s}{8\pi}\,G_{\mu\nu}^aG^{\mu\nu a}\:-\:\sum_{{\rm light}\ q}m_q\bar{q}q\;,
\end{align}
where $B=11-(2/3)N_L$ is the coefficient of the lowest-order term in the QCD beta function, accounting only for the $N_L$ light quarks. The nucleon-nucleon matrix element of the QCD energy-momentum tensor at zero momentum transfer can be written as
\begin{equation}
\braket{N|\big[T^{\rm QCD}\big]_{\mu}^{\phantom{\mu}\mu}|N}\!|_{Q\,=\,0}\ =\ -\:m_N\braket{N| \bar{\psi}_N\psi_N|N}\;,
\end{equation}
where $m_N$ is the nucleon mass. It follows that the effective Higgs-nucleon-nucleon coupling takes the well-known form
\begin{equation}
\mathcal{L}_{\rm eff}\ \supset\ -\:\frac{2\sqrt{2} N_H m_N}{3Bv_{\phi}}\,\bar{\psi}_N\psi_N \phi^0\;,
\end{equation}
wherein the contributions from the $N_L$ light quarks have been omitted. The same result, up to order $\alpha_s(m_{\rm dec})$ corrections to the QCD beta function (where $m_{\rm dec}$ is the scale at which the last heavy quark is decoupled), can be obtained by means of the well-known low energy theorems (see, e.g., Ref.~\cite{Gunion:1989we}).

In terms of the approximate mass eigenstates of the toy model described at the beginning of this section, we have
\begin{equation}
\phi^0\ =\ \frac{h}{\sqrt{2}}\:+\:\frac{2\mu^2}{m_{\phi}^2}\,\frac{v_{\phi}}{\sqrt{2}}\begin{Bmatrix} 1 \\ \frac{v_{\chi}}{M}\end{Bmatrix}\frac{\zeta}{M}\;,
\end{equation}
where $h$ is the SM Higgs boson and $\zeta$ is the light mode that mediates the long-range fifth force. Here, we have accounted for the additonal factor of $\sqrt{2}$ in the normalization of the SM Higgs vev relative to the toy example appearing elsewhere in this work. The two cases in braces correspond to the chameleon and symmetron examples. We therefore find that the fifth force coupling is
\begin{equation}
\mathcal{L}_{\rm eff}\ \supset\ -\:\frac{2N_H}{3B}\,m_N\,\frac{2\mu^2}{m_{\phi}^2}\begin{Bmatrix} 1 \\ \frac{v_{\chi}}{M}\end{Bmatrix}\bar{\psi}_N\psi_N \frac{\zeta}{M}\;.
\end{equation}
As per the arguments in Subsecs.~\ref{sec:chamexamp} and \ref{sec:symexamp}, this vanishes when $\mu^2\to 0$, such that it remains the case that the strength of the fifth force between baryonic matter is modulated by the relative amount of explicit scale breaking in the Higgs sector.

One uncertainty on the coupling to nucleons comes from the contribution of the strange quark. For the light quarks, the nucleon matrix element can be written in the form
\begin{equation}
\label{eq:lightquarkelement}
\braket{N|m_q\bar{q}q|N}\ =\ m_Nf^{N}_{Tq}\;,\qquad q\ \in\ \{{\rm u,d,s}\}\;,
\end{equation}
where the nucleon parameters $f^{N}_{Tq}$ --- the fraction of the nucleon mass carried by the corresponding quarks --- are obtained from measurements of the pion-nucleon sigma term in chiral perturbation theory~\cite{Jungman:1995df,Cheng:1988cz,Cheng:1988im,Gasser:1990ap}. These parameters are commonly taken to be~\cite{Gondolo:2004sc} (see also Ref.~\cite{Cheng:2012qr})
\begin{subequations}
\begin{align}
f_{Tu}^{p}\ =\ 0.023\;,\qquad f_{Td}^{p} \ =\ 0.034\;,\qquad f_{Ts}^{p}\ =\ 0.14\;,\qquad f_{Tc,b,t}^{p}\ =\ 0.0595\;,\\
f_{Tu}^{n}\ =\ 0.019\;,\qquad f_{Td}^{n} \ =\ 0.041\;,\qquad f_{Ts}^{n}\ =\ 0.14\;,\qquad f_{Tc,b,t}^{n}\ =\ 0.0592\;,
\end{align}
\end{subequations}
for protons ($p$) and neutrons ($n$), respectively. In the heavy quark expansion applied here, the fraction of the nucleon mass from gluons is
\begin{equation}
f^N_{TG}\ =\ 1\:-\:\sum_{q\,\in\,\{{\rm u,d,s}\}}f^N_{Tq}\ \approx\ 0.8\;,
\end{equation}
and the heavy-quark matrix elements are given by
\begin{equation}
\braket{N|m_q\bar{q}q|N}\ =\ \frac{2}{27}\,m_Nf^{N}_{TG}\;,\qquad q\ \in\ \{{\rm c,b,t}\}\;.
\end{equation}
After including the additional terms arising from Eq.~\eqref{eq:lightquarkelement}, the contribution of the light quarks to the coupling between nucleons and the conformal scalar is
\begin{equation}
\mathcal{L}_{\rm eff}\ \supset\ -\:m_N\,\frac{2\mu^2}{m_{\phi}^2}\begin{Bmatrix} 1 \\ \frac{v_{\chi}}{M}\end{Bmatrix}\bar{\psi}_N\psi_N\,\frac{\zeta}{M}\,\sum_{q\,\in\,\{{\rm u,d,s}\}}f_{Tq}^N\bigg(1-\frac{2N_H}{3B}\bigg)\;.
\end{equation}

We can parametrize the uncertainty in the coupling by the parameter $\eta$~\cite{Gunion:1989we}, writing
\begin{equation}
\mathcal{L}_{\rm eff}\ \supset\ -\:m_N\,\eta\,\frac{2\mu^2}{m_{\phi}^2}\begin{Bmatrix} 1 \\ \frac{v_{\chi}}{M}\end{Bmatrix}\bar{\psi}_N\psi_N\,\frac{\zeta}{M}\;.
\end{equation}
Neglecting the contributions from the light quarks, we have $\eta=2N_H/(3B)\approx 0.22$ (for $N_H=3$, $B=9$) and $\eta\approx 0.28$ (for $N_H=4$, $B=11-4/3$). Accounting also for the strange quark contribution, we have $\eta=f_{Ts}^N(1-2N_H/(3B))+2N_H/(3B)\approx 0.33$ (i.e.~$N_H=3$, $B=9$). We notice that, in each case, $\eta<1$, such that the coupling strength of the light mode $\zeta$ to the nucleon is parametrically smaller than what one might expect from a Weyl transformation of the nucleon Lagrangian. Any deviation of $\eta$ from unity --- for a recent discussion on determination of the Higgs-nucleon coupling, see Ref.~\cite{Delaunay:2016brc} --- would amount to an effective violation of the weak equivalence principle for hadronic matter versus the elementary fermions, since the strength of the fifth force depends on the details of the binding interactions and does not therefore scale universally with the inertial mass. We leave further discussions of this for future work.


\subsection{Observational bounds}

The discussions above show that if there is an explicit scale-breaking term in the Higgs potential then light conformally coupled scalars will mediate a long-range fifth force. In this subsection, we will show that one can therefore recast Solar System constraints on fifth forces as a constraint on explicit scale breaking in the Higgs sector.

Long-range fifth forces are well constrained experimentally and, for fifth forces mediated by massless fields,\footnote{In fact, we need only that the observations take place on scales smaller than the Compton wavelength of the mediator, such that we can ignore the Yukawa suppression of the fifth force.} the best constraints come from Solar System measurements  \cite{Will:2014kxa,ADELBERGER2009102}.  Within the Solar System, deviations from general relativity can be expressed in the model-independent Parameterised Post Newtonian (PPN) framework. A full description of this framework, and the current constraints on the parameters,  can be found in Ref.~\cite{Will:2014kxa}.  Of interest to us is the $\gamma$ parameter, which determines how much spatial curvature is produced by a unit rest mass. The presence of a fifth force, mediated by a massless field, will appear to observers of a test particle as an extra component of the spatial curvature.  For a fifth force of the form in Eq.~\eqref{eq:usualfifth}, taking the Brans-Dicke/chameleon and toy SM case as an example [cf.~Subsecs.~\ref{sec:massmixing}, \ref{sec:chamexamp} and \ref{sec:symexamp}], we find
\begin{equation}
\label{eq:gammaminusone}
|\gamma-1|\ =\  \bigg|\frac{1}{2+\omega_{\rm eff}}\bigg|\ =\  4\,\bigg(\frac{M_{\rm Pl}^2}{M_{\rm eff}^2}\bigg)\;,
\end{equation}
where, making use of Eq.~\eqref{eq:M2Mpl}, the \emph{effective} Brans-Dicke parameter $\omega_{\rm eff}$ is defined by
\begin{equation}
3+2\omega_{\rm eff}\ \equiv\ \frac{m_{\phi}^4}{4\mu^4}\big(3+2\omega\big)\ =\ \frac{m_{\phi}^4}{4\mu^4}\,\frac{M^2}{2M_{\rm Pl}^2}\;,
\end{equation}
and
\begin{equation}
M_{\rm eff}^2\ \equiv\ \frac{m_{\phi}^4}{4\mu^4}\,M^2\:+\:2M_{\rm Pl}^2\;.
\end{equation}
We note that the factor of $4$ in Eq.~\eqref{eq:gammaminusone} originates from our conventions on the form of the coupling function in Eq.~\eqref{eq:BDcouplingfunc}.

The parameter $\gamma$ is unity in general relativity, and tests of the deflection of light by the Sun and the time delay of signals passing near the Sun, in particular from the tracking of the Cassini satellite,  constrain  $|\gamma -1| <2.3 \times 10^{-5}$ \cite{Bertotti:2003rm}.  This implies that $M_{\rm eff}^2\gtrsim 2\times 10^{5}\, M_{\rm Pl}^2$ and yields the well-known bound $\omega\gtrsim 4\times 10^4$ for $M_{\rm eff}^2=M^2\gg M_{\rm Pl}^2$. However, for fixed $M\neq M_{\rm eff}$, the bound on $\gamma$ can actually be translated into an illustrative bound on the scale-breaking parameter $\mu$ as
\begin{equation}
\frac{\mu}{m_\phi}\ \lesssim\ 0.03\,\bigg(\frac{M}{M_{\rm Pl}}\bigg)^{1/2}\;,
\end{equation}
which can relax the constraint on $\omega\ll\omega_{\rm eff}$. For $M\sim M_{\rm Pl}$, we therefore find that an explicit scale can be responsible for $\lesssim 3\,\%$ of the total mass. Extrapolating this to the SM with a modified Higgs sector, i.e.~taking $m_\phi\approx m_h=125\;{\rm GeV}$ --- and ignoring the hadronic uncertainties on the Higgs-nucleon-nucleon coupling --- we would require $\mu\lesssim 4\ {\rm GeV}$. 

Finally, we note that these bounds assume that any screening mechanisms are inactive within the Solar System. Whilst the details will be more involved, modifications to the origin of the symmetry breaking along the lines described at the beginning of this section would still yield a suppression of the fifth force over that arising from the screening.


\section{Higgs-dilaton}
\label{sec:HiggsDilaton}

Finally, we turn to the case when there are no explicit scale-breaking terms. We focus, in particular, on the Higgs-dilaton theory, which has been studied extensively in the literature~\cite{Wetterich:1987fm,Buchmuller:1988cj,Shaposhnikov:2008xb,Shaposhnikov:2008xi,Blas:2011ac,GarciaBellido:2011de,GarciaBellido:2012zu,Bezrukov:2012hx,Henz:2013oxa,Rubio:2014wta,Karananas:2016grc,Ferreira:2016vsc,Ferreira:2016wem,Ferreira:2016kxi,Ferreira:2018qss}.
This model provides a concrete example in which to show that the kinetic mixing of the scalar fields also does not contribute a fifth force in the scale-invariant limit. In the appendices, and for completeness, we show how the absence of fifth forces is borne out at linear order in fluctuations, both in the Einstein and Jordan frames.

The Higgs-dilaton model extends the SM with a singlet scalar field, as we have done in the preceding sections, but it includes non-minimal couplings of both the singlet scalar and the Higgs field to gravity. The specific way in which this is done, as well as the specific choice of symmety-breaking potential, mean that the  Higgs-dilaton model is a realization of a no-scale scenario, wherein the scalar, and an associated spontaneous breaking of scale symmetry, are responsible for generating \emph{all} other scales.  The interactions between the singlet scalar and Higgs field induce electroweak symmetry breaking, explaining the masses of the electroweak gauge bosons via the standard Higgs mechanism, the Yukawa interactions of the Higgs field give rise to the fermion masses, and the non-minimal gravitational couplings of the scalar fields generate the Planck scale. Moreover, by supplementing the SM with right-handed singlet neutrinos, Majorana mass terms can be generated by the singlet scalar directly. An example of such a scenario is the embedding of the Higgs-dilaton model in the $\nu$MSM~\cite{Asaka:2005pn,Asaka:2005an}, which has the Lagrangian
\begin{align}
\mathcal{L}_{\nu{\rm MSM}}\ &=\ \mathcal{L}_{{\rm SM,\,V(\phi)\,\to\,0}}\:+\:\mathcal{L}_G\:-\:\frac{1}{2}\,g^{\mu\nu}\partial_{\mu}\chi\partial_{\nu}\chi\:-\:U(\phi,\chi)\:\nonumber\\&-\:\Big(\frac{1}{2}\,\bar{N}_iig^{\mu\nu}\gamma_{\mu}\overset{\leftrightarrow}{\partial}_{\nu}N_i\:+\:\frac{1}{2}\,f_{ij}\bar{N}^c_i\chi N_j\:+\:y^{(N)}_{\alpha i}\bar{L}_{\alpha}\tilde{\phi} N_i\:\:+\:{\rm H.c.}\Big)\;,
\end{align}
where $\mathcal{L}_{{\rm SM,V(\phi)\,\to\,0}}$ is the SM Lagrangian less the SM Higgs potential $V(\phi)$, $\mathcal{L}_G$ is the Lagrangian of the gravity sector
\begin{equation}
\mathcal{L}_G\ = \ \frac{1}{2}\Big(2\,\xi_{\phi}\phi^{\dag}\phi\:+\:\xi_{\chi}\chi^2\Big) \mathcal{R}\;,
\end{equation}
and the scalar-field potential is
\begin{equation}
U(\phi,\chi)\ =\ \lambda\bigg(\phi^{\dag}\phi\:-\:\frac{\beta}{2\lambda}\chi^2\bigg)^{\!2}\;.
\end{equation}
Here, $\tilde{\phi}=i\sigma^2\phi^*$ is the isospin conjugate of the SM Higgs doublet $\phi$ (where $\sigma^2$ is the second Pauli matrix), $\chi$ is the singlet scalar field, the $L_{\alpha}$ ($\alpha=e,\mu,\tau$) are the SM lepton doublets, the $N_i$ ($i=1,2,3$) are three generations of right-handed singlet leptons, the superscript $c$ denotes charge conjugation, and $\mathcal{R}$ is the Ricci scalar, as before. The $f_{ij}$ are the singlet-scalar Yukawa couplings responsible for the Majorana mass matrix, and the \smash{$y^{(N)}_{\alpha i}$} are Higgs Yukawa couplings responsible for the Dirac neutrino mass matrix, where the superscript $N$ is to differentiate these couplings from those that yield the Dirac masses of the electrically-charged SM leptons, i.e.~\smash{$y^{(L)}_{\alpha\beta}\bar{L}_{\alpha}\,\phi\, e_{\beta}+{\rm H.c.}$}. Note that all of the Yukawa couplings are matrices in flavour space.

In order to avoid the technical complications that arise from the gauge and flavour structure of the $\nu$MSM, we will study a simplified model that comprises the real prototype of the Higgs mechanism. Our toy model has the Jordan-frame action
\begin{equation}\label{Jordan-Frame-Action}
S\ =\ \int{\rm d}^4x\;\sqrt{-\,g}\,\bigg[\frac{1}{2}\,F(\phi,\chi)\mathcal{R}\:+\:\,\mathcal{L}_{\phi,\chi}\:+\:\mathcal{L}_{\psi,N}\bigg]\;,
\end{equation}
where the non-minimal coupling function is
\begin{equation}
\label{eq:Fdef}
F(\phi,\chi)\ =\ \xi_{\phi}\,\phi^2\:+\:\xi_{\chi}\,\chi^2\;,
\end{equation}
and the scalar sector has Lagrangian
\begin{equation}
\mathcal{L}_{\phi,\chi}\ =\ -\:\frac{1}{2}\,g^{\mu\nu}\,\partial_{\mu}\phi\,\partial_{\nu}\phi\:-\:\frac{1}{2}\,g^{\mu\nu}\,\partial_{\mu}\chi\partial_{\nu}\chi\:-\:U(\phi,\chi)\;,
\end{equation}
with potential
\begin{equation}
\label{eq:HDpot}
U(\phi,\chi)\ =\ \frac{\lambda}{4}\,\bigg(\phi^2\:-\:\frac{\beta}{\lambda}\,\chi^2\bigg)^{\!2}\;.
\end{equation}
As proxies for the SM fermions and the right-handed neutrinos, we take two fermion fields $\psi$ and $N$, whose (Dirac) masses are obtained through Yukawa couplings to the would-be Higgs field $\phi$ and the would-be singlet scalar $\chi$, respectively:
\begin{equation}
\mathcal{L}_{\psi,N}\ =\ -\:\bar{\psi}i\overset{\leftrightarrow}{\tilde{\slashed{\partial}}}
\psi\:-\:\bar{N}i\overset{\leftrightarrow}{\tilde{\slashed{\partial}}}
N\:-\:\bar{\psi}\phi\psi\:-\:\bar{N}\chi N\:-\:\bar{\psi}\phi N\:-\:\bar{N}\phi \psi\;.
\end{equation}
We have set all of the Yukawa couplings to unity for convenience, and the term $\bar{N}\phi N$, which is permitted for this toy model, has been precluded, so as to emulate the interactions of the $\nu$MSM. For arbitrary $\xi_{\phi,\chi}$, we will see that, while the fermion sector is locally Weyl invariant, the Higgs and singlet scalar kinetic terms are only scale, viz.~globally Weyl invariant.

The first step in determining the presence, or absence, of a fifth force in this theory is to determine whether the scalar fields have non-trivial field profiles sourced by finite configurations of the fermion fields. We find that the Klein-Gordon and Dirac equations take the forms
\begin{subequations}
\label{eq:system}
\begin{gather}
\label{eq:HiggsKG}
\Box\, \phi\:+\:\xi_{\phi} \mathcal{R}\,\phi\:-\:\lambda\,\phi\,\Big(\phi^2-\frac{\beta}{\lambda}\,\chi^2\Big)\:-\:\bar{\psi}\psi\:-\:\bar{\psi}N\:-\:\bar{N}\psi\ =\ 0\;,\\
\label{eq:chiKG}
\Box\,\chi\:+\:\xi_{\chi} \mathcal{R}\,\chi\:+\:\beta\,\chi\,\Big(\phi^2-\frac{\beta}{\lambda}\,\chi^2\Big)\:-\:\bar{N} N\ =\ 0\;,\\
-\:i\slashed{\nabla}\psi\:-\:\phi\psi\:-\:\phi N\ =\ 0\;,\\
-\:i\slashed{\nabla}N\:-\:\chi N\:-\:\phi \psi\ =\ 0\;,
\end{gather}
\end{subequations}
and the Einstein equations can be written as
\begin{align}
F(\phi,\chi)G_{\mu\nu}\ &=\ T_{\mu\nu}\:+\:\partial_{\mu}\phi\,\partial_{\nu}\phi\:+\:\partial_{\mu}\chi\,\partial_{\nu}\chi\:-\:\frac{1}{2}\,g_{\mu\nu}\,\Big(\partial_{\rho}\phi\,\partial^{\rho}\phi\:+\:\partial_{\rho}\chi\,\partial^{\rho}\chi\Big)\nonumber\\&\qquad+\:\nabla_{\mu}\,\partial_{\nu}\,F(\phi,\chi)\:-\:g_{\mu\nu}\,\Box\,F(\phi,\chi)\:-\:g_{\mu\nu}\,U(\phi,\chi)\;,
\end{align}
where $G_{\mu\nu}=\mathcal{R}_{\mu\nu}-g_{\mu\nu}\mathcal{R}/2$ and
\begin{align}
T_{\mu\nu}\ &=\ -\:\frac{2}{\sqrt{-\,g}}\,\frac{\delta \sqrt{-\,g}\,\mathcal{L}_{\psi,N}}{\delta g^{\mu\nu}}\nonumber \\&=\ \frac{1}{2}\,\bar{\psi}i\gamma_{(\mu}\nabla_{\nu)}\psi\:+\:\frac{1}{2}\,\bar{N}i\gamma_{(\mu}\nabla_{\nu)}N\nonumber\\&\qquad-\:g_{\mu\nu}\Big(\bar{\psi}i\slashed{\nabla}\psi\:+\:\bar{\psi}\phi\psi\:+\:\bar{N}i\slashed{\nabla}N\:+\:\bar{N}\chi N\:+\:\bar{\psi}\phi N\:+\:\bar{N}\phi\psi\Big)
\label{eq:energymomentum}
\end{align}
is the energy-momentum tensor of the matter fields. Notice that we have symmetrized the Lorentz indices of the kinetic term, i.e.~$\gamma_{(\mu}\nabla_{\nu}) \equiv \gamma_{\mu}\nabla_{\nu}+\gamma_{\nu}\nabla_{\mu}$, where $\nabla_{\mu}\equiv \partial_{\mu}-\frac{i}{4}\omega_{\nu\rho\mu}\sigma^{\nu\rho}$ is the covariant derivative in which $\omega_{\nu\rho\mu}=e_{\nu a}(\partial_{\mu}e^a_{\rho}+\Gamma^{a}_{bc}e^b_{\rho}e^c_{\mu})$ is the spin connection and $\sigma^{\mu\nu}=\frac{i}{2}[\gamma^{\mu},\gamma^{\nu}]$. In four dimensions, the trace of the energy momentum tensor is given by
\begin{equation}
T_{\mu}^{\phantom{\mu}\mu}\ =\ -\:3\,\bar{\psi}i\slashed{\nabla}\psi\:-\:3\,\bar{N}i\slashed{\nabla}N\:-\:\,4\Big(\bar{\psi}\phi\psi\:+\:\bar{N}\chi N\:+\:\bar{\psi}\phi N\:+\:\bar{N}\phi\psi\Big)\;,
\end{equation}
which reduces to
\begin{equation}
T_{\mu}^{\phantom{\mu}\mu}\big|_{\rm OS}\ =\ -\:\bar{\psi}\phi\psi\:-\:\bar{N}\chi N\:-\:\bar{\psi}\phi N\:-\:\bar{N}\phi \psi
\end{equation}
on-shell.

Taking the trace of the Einstein equations and evaluating on-shell, we arrive at
\begin{equation}
\label{eq:traceEinstein}
g_{\phi}\,\Box\,\phi^2\:+\:g_{\chi}\,\Box\,\chi^2\ =\ 0\;,
\end{equation}
where we have defined
\begin{equation}
g_{\phi,\chi}\ \equiv\ 6\,\xi_{\phi,\chi}\:+\:1\;.
\end{equation}
From Eq.~\eqref{eq:traceEinstein}, we immediately see two things: firstly, there exists a massless mode in this theory
\begin{equation}
\label{eq:Dilatondef}
\sigma\ \equiv \ \frac{M}{2}\ln \left(\big(g_{\phi}\,\phi^2+g_{\chi}\,\chi^2)/M^2\right)\;,
\end{equation}
which we will call the dilaton; secondly, this massless mode is not sourced by any of the fermion fields. As the dilaton cannot be sourced by matter density, it does not yield potentially dangerous fifth forces.  The scale $M$ introduced here plays the same role as the cut-off scale in the preceding sections.

In fact, Eq.~\eqref{eq:traceEinstein} is nothing other than the conservation law for the dilatation current. Notice that the left-hand side vanishes identically in the conformal limit $\xi_{\phi,\chi}\to-\,1/6$. The dilaton is the Goldstone boson of the spontaneously broken scale symmetry. Making the Weyl rescaling of the metric
\begin{equation}
\check{g}_{\mu\nu}\ \equiv\ e^{2\sigma/M}g_{\mu\nu}\;,
\end{equation}
the equation of motion for the dilaton can be written in the simple form
\begin{equation}
\label{eq:dilatoneom}
\check{\Box}\,\sigma\ =\ 0\;.
\end{equation}

The dilaton is protected by a shift symmetry. It can therefore have at most derivative couplings to the Higgs field. In order to study these couplings, it is convenient to move to the Einstein frame. We do this through two consecutive Weyl rescalings. While this could, of course, be combined into a single transformation [cf.~App.~\ref{sec:linearE}], this two-step procedure is more illustrative, allowing us to isolate the dilaton more easily.

Following Ref.~\cite{Brax:2014baa}, we first scale out the dilaton by making the field redefinitions
\begin{equation}
\check{\phi} \equiv\ e^{-\sigma/M}\phi\;,\qquad \check{\chi}\ \equiv\ e^{-\sigma/M}\chi\;,
\end{equation}
such that
\begin{equation}
\label{eq:fphichi}
F(\phi,\chi)\ =\ e^{2\sigma/M}F(\check{\phi},\check{\chi})\;.
\end{equation}
We now rescale the metric, defining
\begin{equation}
\check{g}_{\mu\nu}\ \equiv\ e^{2\sigma/M}g_{\mu\nu}\;.
\end{equation}
The Ricci scalars of the two metrics are related by
\begin{equation}
\mathcal{R}\ =\ e^{2\sigma/M}\Big(\check{\mathcal{R}}\:+\:6\,\check{g}^{\mu\nu}\,\check{\nabla}_{\mu}\partial_{\nu}\frac{\sigma}{M}\:-\:6\,\check{g}^{\mu\nu}\,\partial_{\mu}\frac{\sigma}{M}\,\partial_{\nu}\frac{\sigma}{M}\Big)\;,
\label{eq:riccis}
\end{equation}
and the action in Eq.~\eqref{Jordan-Frame-Action} becomes
\begin{align}\label{Jordan-Frame-Action1}
S\ &=\ \int\!{\rm d}^4x\;\sqrt{-\,\check{g}}\;\bigg[\frac{1}{2}\,F(\check{\phi},\check{\chi})\,\check{\mathcal{R}}\:-\:3\,\check{g}^{\mu\nu}\partial_{\mu}F(\check{\phi},\check{\chi})\,\partial_{\nu}\frac{\sigma}{M}\:-\:3\,\check{g}^{\mu\nu}\,F(\check{\phi},\check{\chi})\,\partial_{\mu}\frac{\sigma}{M}\,\partial_{\nu}\frac{\sigma}{M}\nonumber\\&\qquad-\:\frac{1}{2}\,\check{g}^{\mu\nu}\,e^{-2\sigma/M}\,\partial_{\mu}(e^{\sigma/M}\check{\phi})\,\partial_{\nu}(e^{\sigma/M}\check{\phi})\:-\:\frac{1}{2}\,\check{g}^{\mu\nu}\,e^{-2\sigma/M}\,\partial_{\mu}(e^{\sigma/M}\check{\chi})\,\partial_{\nu}(e^{\sigma/M}\check{\chi})\nonumber\\&\qquad-\:U(\check{\phi},\check{\chi})\:-\:\bar{\check{\psi}}i\overset{\leftrightarrow}{\check{\slashed{\partial}}}\check{\psi}\:-\:\check{N}i\overset{\leftrightarrow}{\check{\slashed{\partial}}}\check{N}\:-\:\bar{\check{\psi}}\check{\phi}\check{\psi}\:-\:\bar{\check{N}}\check{\chi} \check{N}\:-\:\bar{\check{\psi}}\check{\phi}\check{N}\:-\:\bar{\check{N}}\check{\phi} \check{\psi}\bigg]\;,
\end{align}
where we have also  rescaled the fermion fields
\begin{equation}
\check{\psi}\ \equiv\ e^{-(3/2)\sigma/M}\psi\;,\qquad \check{N}\ \equiv\ e^{-(3/2)\sigma/M}N\;.
\end{equation}
At this point it appears that we have three scalar fields $\sigma$, $\check{\phi}$ and $\check{\chi}$. We must remember, however, that there is a constraint equation [see Eq.~\eqref{eq:constraint} below], which renders one of these scalars non-dynamical. 

We now perform a second Weyl rescaling, defining
\begin{equation}
\tilde{g}_{\mu\nu}\ =\ \frac{F(\check{\phi},\check{\chi})}{M_{\rm Pl}^2}\,\check{g}_{\mu\nu}\;.
\end{equation}
As in Eq.~\eqref{eq:riccis}, the Ricci scalars are related by
\begin{equation}
\label{eq:RtoRtilde}
\check{\mathcal{R}}\ =\ \frac{F(\check{\phi},\check{\chi})}{M_{\rm Pl}^2}\bigg[\tilde{\mathcal{R}}\:+\:3\,\tilde{g}^{\mu\nu}\,\tilde{\nabla}_{\mu}\partial_{\nu}\,\ln F(\check{\phi},\check{\chi})\:-\:\frac{3}{2}\,\tilde{g}^{\mu\nu}\partial_{\mu}\ln F(\check{\phi},\check{\chi})\,\partial_{\nu}\ln F(\check{\phi},\check{\chi})\bigg]\;,
\end{equation}
and the action in Eq.~\eqref{Jordan-Frame-Action1} can be written as
\begin{align}
S\ &=\ \int\!{\rm d}^4x\;\sqrt{-\,\tilde{g}}\;\bigg[\frac{1}{2}\,M_{\rm Pl}^2\,\tilde{\mathcal{R}}\:-\:\frac{3}{4}\,\tilde{g}^{\mu\nu}\,M_{\rm Pl}^2\,\partial_{\mu}\ln\frac{F(\check{\phi},\check{\chi})}{M_{\rm Pl}^2}\,\partial_{\nu}\ln\frac{F(\check{\phi},\check{\chi})}{M_{\rm Pl}^2}\nonumber\\&\qquad -\:3\,\tilde{g}^{\mu\nu}\,M_{\rm Pl}^2\,\partial_{\mu}\ln\frac{F(\check{\phi},\check{\chi})}{M_{\rm Pl}^2}\,\partial_{\nu}\frac{\sigma}{M}\:-\:3\,\tilde{g}^{\mu\nu}\,M_{\rm Pl}^2\,\partial_{\mu}\frac{\sigma}{M}\,\partial_{\nu}\frac{\sigma}{M}\nonumber\\&\qquad-\:\frac{1}{2}\,\tilde{g}^{\mu\nu}\,\frac{M_{\rm Pl}^2\,e^{-2\sigma/M}}{F(\check{\phi},\check{\chi})}\,\partial_{\mu}(e^{\sigma/M}\check{\phi})\,\partial_{\nu}(e^{\sigma/M}\check{\phi})\nonumber\\&\qquad-\:\frac{1}{2}\,\tilde{g}^{\mu\nu}\,\frac{M_{\rm Pl}^2\,e^{-2\sigma/M}}{F(\check{\phi},\check{\chi})}\,\partial_{\mu}(e^{\sigma/M}\check{\chi})\,\partial_{\nu}(e^{\sigma/M}\check{\chi})\nonumber\\&\qquad -\:\frac{M_{\rm Pl}^4}{F^2(\check{\phi},\check{\chi})}\,U(\check{\phi},\check{\chi})\:-\:\bar{\tilde{\psi}}i\overset{\leftrightarrow}{\tilde{\slashed{\partial}}}\tilde{\psi}\:-\:\bar{\tilde{N}}i\overset{\leftrightarrow}{\tilde{\slashed{\partial}}}\tilde{N}\nonumber\\&\qquad-\:\frac{M_{\rm Pl}}{F^{1/2}(\check{\phi},\check{\chi})}\Big(\bar{\tilde{\psi}}\check{\phi}\tilde{\psi}\:+\:\bar{\tilde{N}}\check{\chi} \tilde{N}\:+\:\bar{\tilde{\psi}}\check{\phi}\tilde{N}\:+\:\bar{\tilde{N}}\check{\phi} \tilde{\psi}\Big)\bigg]\;,
\label{eq:efaction}
\end{align}
where we have performed a final rescaling of the fermion fields, defining
\begin{equation}
\tilde{\psi}\ \equiv\ \frac{M_{\rm Pl}^{3/2}}{F^{3/4}(\check{\phi},\check{\chi})}\,\check{\psi}\;,\qquad \tilde{N}\ \equiv\ \frac{M_{\rm Pl}^{3/2}}{F^{3/4}(\check{\phi},\check{\chi})}\,\check{N}\;.
\end{equation}

After some algebra, the scalar and gravitational parts of the action in Eq.~\eqref{eq:efaction} can be rewritten in the form
\begin{align}
S\ &\supset\ \int\!{\rm d}^4x\;\sqrt{-\,\tilde{g}}\;\bigg[\frac{1}{2}\,M_{\rm Pl}^2\,\tilde{\mathcal{R}}\:-\:\frac{3}{4}\,\tilde{g}^{\mu\nu}\,M_{\rm Pl}^2\,\partial_{\mu}\ln\frac{F(\check{\phi},\check{\chi})}{M_{\rm Pl}^2}\,\partial_{\nu}\ln\frac{F(\check{\phi},\check{\chi})}{M_{\rm Pl}^2}\nonumber\\&\qquad -\:\frac{1}{2}\,\tilde{g}^{\mu\nu}\,\frac{M_{\rm Pl}^2}{F(\check{\phi},\check{\chi})}\,\partial_{\mu}\big(g_{\phi}\,\check{\phi}^2+g_{\chi}\,\check{\chi}^2\big)\,\partial_{\nu}\frac{\sigma}{M}\nonumber\\&\qquad-\:\frac{1}{2}\,\tilde{g}^{\mu\nu}\,\frac{M_{\rm Pl}^2}{F(\check{\phi},\check{\chi})}\,\big(g_{\phi}\,\check{\phi}^2+g_{\chi}\,\check{\chi}^2\big)\,\partial_{\mu}\frac{\sigma}{M}\,\partial_{\nu}\frac{\sigma}{M}\nonumber\\&\qquad-\:\frac{1}{2}\,\tilde{g}^{\mu\nu}\,\frac{M_{\rm Pl}^2}{F(\check{\phi},\check{\chi})}\,\partial_{\mu}\check{\phi}\,\partial_{\nu}\check{\phi}\:-\:\frac{1}{2}\,\tilde{g}^{\mu\nu}\,\frac{M_{\rm Pl}^2}{F(\check{\phi},\check{\chi})}\,\partial_{\mu}\check{\chi}\,\partial_{\nu}\check{\chi}\nonumber\\&\qquad -\:\frac{M_{\rm Pl}^4}{F^2(\check{\phi},\check{\chi})}\,U(\check{\phi},\check{\chi})\bigg]\;.
\end{align}
Choosing the constraint --- which is really just a choice of normalization for the massive degree of freedom --- to be
\begin{equation}
\label{eq:constraint}
g_{\phi}\,\check{\phi}^2\:+\:g_{\chi}\,\check{\chi}^2\ =\ M^2\;,
\end{equation}
the scalar and gravitational parts of the action become
\begin{align}
S\ &\supset\ \int\!{\rm d}^4x\;\sqrt{-\,\tilde{g}}\;\bigg[\frac{1}{2}\,M_{\rm Pl}^2\,\tilde{\mathcal{R}}\:-\:\frac{3}{4}\,\tilde{g}^{\mu\nu}\,M_{\rm Pl}^2\,\partial_{\mu}\ln\frac{F(\check{\phi},\check{\chi})}{M_{\rm Pl}^2}\,\partial_{\nu}\ln\frac{F(\check{\phi},\check{\chi})}{M_{\rm Pl}^2}\nonumber\\&\qquad-\:\frac{1}{2}\,\tilde{g}^{\mu\nu}\,\frac{M_{\rm Pl}^2}{F(\check{\phi},\check{\chi})}\,\partial_{\mu}\sigma\,\partial_{\nu}\sigma\:-\:\frac{1}{2}\,\tilde{g}^{\mu\nu}\,\frac{M_{\rm Pl}^2}{F(\check{\phi},\check{\chi})}\,\partial_{\mu}\check{\phi}\,\partial_{\nu}\check{\phi}\:-\:\frac{1}{2}\,\tilde{g}^{\mu\nu}\,\frac{M_{\rm Pl}^2}{F(\check{\phi},\check{\chi})}\,\partial_{\mu}\check{\chi}\,\partial_{\nu}\check{\chi}\nonumber\\&\qquad -\:\frac{M_{\rm Pl}^4}{F^2(\check{\phi},\check{\chi})}\,U(\check{\phi},\check{\chi})\bigg]\;.
\end{align}
Note that this choice of constraint is consistent with the definition of the dilaton in Eq.~\eqref{eq:Dilatondef}.
Varying with respect to $\sigma$, we obtain the equation of motion
\begin{equation}
\label{eq:Dilatonfulleq}
\frac{1}{\sqrt{-\,\tilde{g}}}\,\partial_{\mu}\Big(\sqrt{-\,\tilde{g}}\,\tilde{g}^{\mu\nu}F^{-1}(\check{\phi},\check{\chi})\,\partial_{\nu}\sigma\Big)\ =\ 0\;,
\end{equation}
cf.~Refs.~\cite{Brax:2014baa} and~\cite{Ferreira:2016kxi}. At the background level, we can take $\braket{\sigma}=0$, such that $\braket{F(\check{\phi},\check{\chi})}=\braket{F(\phi,\chi)}=M_{\rm Pl}^2$, and it is clear that Eq.~\eqref{eq:Dilatonfulleq} is consistent with Eq.~\eqref{eq:dilatoneom}. 

We can proceed now to eliminate any remaining dependence on $\check{\chi}$. To this end, we quote the following results:
\begin{subequations}
\label{eq:identities}
\begin{gather}
\check{\chi}^2\ =\ \frac{M^2}{g_{\chi}}\bigg(1\:-\:\frac{g_{\phi}\,\check{\phi}^2}{M^2}\bigg)\;,\\
\partial_{\mu}\check{\chi}\,\partial_{\nu}\check{\chi}\ =\ \frac{g_{\phi}}{g_{\chi}}\,\frac{g_{\phi}\,\check{\phi}^2}{M^2}\bigg(1\:-\:\frac{g_{\phi}\,\check{\phi}^2}{M^2}\bigg)^{-1}\partial_{\mu}\check{\phi}\,\partial_{\nu}\check{\phi}\;,\\
\partial_{\mu}\ln\frac{F(\check{\phi},\check{\chi})}{M_{\rm Pl}^2}\,\partial_{\nu}\ln\frac{F(\check{\phi},\check{\chi})}{M_{\rm Pl}^2}\ =\ \frac{1}{9}\,\frac{M_{\rm Pl}^2}{F^2(\check{\phi},\check{\chi})}\bigg(1\:-\:\frac{g_{\phi}}{g_{\chi}}\bigg)^{\!2}\check{\phi}^2\,\partial_{\mu}\check{\phi}\,\partial_{\nu}\check{\phi}\;.
\end{gather}
\end{subequations}
We also note that
\begin{equation}
F(\check{\phi},\check{\chi})\ =\ \xi_{\phi}\,\check{\phi}^2\:+\:\frac{\xi_{\chi}M^2}{g_{\chi}}\bigg(1\:-\:\frac{g_{\phi}\,\check{\phi}^2}{M^2}\bigg)\;,
\end{equation}
although we will not employ this directly. Making use of Eq.~\eqref{eq:identities}, the action can be written in the final form
\begin{align}\label{Einstein-Frame-tilde}
S\ &=\ \int\!{\rm d}^4x\;\sqrt{-\,\tilde{g}}\;\bigg\{\frac{1}{2}\,M_{\rm Pl}^2\,\tilde{\mathcal{R}}\:-\:\frac{1}{2}\,\tilde{g}^{\mu\nu}\,\frac{M_{\rm Pl}^2}{F(\check{\phi},\check{\chi})}\,\partial_{\mu}\sigma\,\partial_{\nu}\sigma\nonumber\\&\quad-\:\frac{1}{2}\,\tilde{g}^{\mu\nu}\,\frac{M_{\rm Pl}^2}{F(\check{\phi},\check{\chi})}\bigg[1\:+\:\frac{1}{6}\,\frac{\check{\phi}^2}{F(\check{\phi},\check{\chi})}\,\bigg(1\:-\:\frac{g_{\phi}}{g_{\chi}}\bigg)^{\!2}\:+\:\frac{g_{\phi}}{g_{\chi}}\,\frac{g_{\phi}\,\check{\phi}^2}{M^2}\bigg(1\:-\:\frac{g_{\phi}\,\check{\phi}^2}{M^2}\bigg)^{\!-1}\;\bigg]\,\partial_{\mu}\check{\phi}\,\partial_{\nu}\check{\phi}\nonumber\\&\quad -\:\frac{M_{\rm Pl}^4}{F^2(\check{\phi},\check{\chi})}\,U(\check{\phi},\check{\chi})-\:\bar{\tilde{\psi}}i\overset{\leftrightarrow}{\tilde{\slashed{\partial}}}\tilde{\psi}\:-\:\bar{\tilde{N}}i\overset{\leftrightarrow}{\tilde{\slashed{\partial}}}\tilde{N}\nonumber\\&\quad-\:\frac{M_{\rm Pl}}{F^{1/2}(\check{\phi},\check{\chi})}\Big(\bar{\tilde{\psi}}\check{\phi}\tilde{\psi}\:+\:\frac{M}{\sqrt{g_{\chi}}}\,\bar{\tilde{N}}\bigg(1\:-\:\frac{g_{\phi}\,\check{\phi}^2}{M^2}\bigg)^{\!1/2} \tilde{N}\:+\:\bar{\tilde{\psi}}\check{\phi}\tilde{N}\:+\:\bar{\tilde{N}}\check{\phi} \tilde{\psi}\Big)\bigg\}\;,
\end{align}
with potential
\begin{equation}
U(\check{\phi},\check{\chi})\ =\ \frac{\tilde{\lambda}}{4}\,\big(\check{\phi}^2-\tilde{v}_{\phi}^2\big)^2\;,
\end{equation}
where
\begin{equation}
\tilde{\lambda}\ \equiv\ \lambda\bigg(1\:+\:\frac{\beta}{\lambda}\,\frac{g_{\phi}}{g_{\chi}}\bigg)^{\!2}\;,\qquad \tilde{v}_{\phi}^2\ \equiv\ \frac{\beta}{\lambda}\,\frac{M^2}{g_{\chi}}\,\bigg(1\:+\:\frac{\beta}{\lambda}\frac{g_{\phi}}{g_{\chi}}\bigg)^{\!-1}\;.
\end{equation}

By expanding this action around the vacuum expectation values in the broken phase, we can determine the nature of the mixing between the Higgs and the dilaton at leading order. We write the Higgs field as $\check{\phi} =  \tilde{v}_{\phi} + \tilde{h}$. Realizing that
\begin{equation}
F(\check{\phi},\check{\chi})\ =\ M_{\rm Pl}^2\:+\:2\,\frac{\xi_{\phi}-\xi_{\chi}}{g_{\chi}}\,\tilde{v}_{\phi}\,\tilde{h}\:+\:\mathcal{O}(\tilde{h}^2)\;,
\end{equation}
the scalar part of the action, at quadratic order in the fluctuations, becomes
\begin{align}\label{Einstein-Frame-h}
S\ &\supset\ \int\!{\rm d}^4x\;\sqrt{-\,\tilde{g}}\;\bigg[\frac{1}{2}\,M_{\rm Pl}^2\,\tilde{\mathcal{R}}\:-\:\frac{1}{2}\,\tilde{g}^{\mu\nu}\,\partial_{\mu}\sigma\,\partial_{\nu}\sigma\:-\:\frac{1}{2}\,\tilde{g}^{\mu\nu}\,\partial_{\mu}\tilde{h}\,\partial_{\nu}\tilde{h}\:-\:\frac{1}{2}\,m_h^2\,\tilde{h}^2\:+\:\cdots\bigg]\;.
\end{align}
where $m_h^2=2\tilde{\lambda}\tilde{v}_{\phi}^2$ and we have omitted subdominant terms in $\tilde{v}_{\phi}/M\ll 1$ and $\tilde{v}_{\phi}/M_{\rm Pl}\ll 1$.

We see from Eqs.~\eqref{Einstein-Frame-tilde} and~\eqref{Einstein-Frame-h} that, at quadratic order in the fluctuations, the dilaton does not couple to the fermions either directly or indirectly. Moreover, we see that there is no kinetic mixing between the Higgs field and the dilaton at quadratic order, despite such a mixing being permitted by the dilaton shift symmetry. The absence of this kinetic mixing is consistent with our earlier observations [cf.~Eq.~\eqref{eq:Yukkin} and the discussion in Sec.~\ref{sec:ExplicitScale}] for the fully scale-invariant case. Hence, and as a result of the dilatation symmetry of this model, we see explicitly that the dilaton cannot give rise to long-range fifth forces in agreement with Refs.~\cite{Shaposhnikov:2008xb,Brax:2014baa,Ferreira:2016kxi}.

Finally, we remark that, at third order in the fluctuations, there is a derivative interaction between the Higgs and the dilaton:
\begin{equation}
\mathcal{L}\ \supset\ \frac{\xi_{\phi}-\xi_{\chi}}{g_{\chi}}\,\frac{\tilde{v}_{\phi}}{M_{\rm Pl}^2}\,\tilde{h}\,\tilde{g}^{\mu\nu}\,\partial_{\mu}\sigma\,\partial_{\nu}\sigma\;.
\end{equation}
Not only is this term Planck-suppressed, but it only contributes at most a loop-level correction to the Higgs propagator. Namely, this is a self-energy correction of the form
\begin{equation}
i\Pi(p^2)\ \supset\ i^2\bigg(\frac{\xi_{\phi}-\xi_{\chi}}{g_{\chi}}\bigg)^{\!2}\,\bigg(\frac{\tilde{v}_{\phi}}{M_{\rm Pl}^2}\bigg)^{\!2}\int\!\frac{{\rm d}^4k}{(2\pi)^4}\;\frac{k^2(p-k)^2}{[k^2-i\epsilon][(p-k)^2-i\epsilon]}\;,
\end{equation}
which is, in fact, zero in dimensional regularization. Hence, there can be no fifth force introduced also by this derivative interaction.



\section{Conclusions}
\label{sec:Conclusions}

We have illustrated how the presence of explicit scale-breaking terms in the SM impacts upon scalar fifth forces in scalar-tensor modifications of gravity involving non-minimal gravitational couplings. In so doing, we have shown that these particular modifications of general relativity are equivalent to Higgs-portal theories and that their fifth-force phenomenology depends strongly on the structure of the SM. As a result, we have argued that the non-observation of fifth forces can be interpreted as a constraint on the structure of the SM Higgs sector and the origin of its symmetry breaking, providing an upper bound on any explicit scale-breaking term.  In other words, if one assumes that light, non-minimally coupled scalar fields exist in Nature, Solar System tests of gravity can, quite remarkably, tell us about the structure of the SM and the origin of its symmetry breakings.  The import of this final observation is that our understanding of fifth forces and the behaviour of modified theories of gravity rests on our knowledge of how scales emerge in the SM and its extensions.


\begin{acknowledgments}
This work was supported by STFC Grant No.~ST/L000393/1, a Leverhulme Trust Research Leadership Award and a Royal Society University Research Fellowship. The authors would like to thank Philippe Brax, Eugenio Del Nobile, Pedro Ferreira, Bj\"{o}rn Garbrecht, Carlos Tamarit and Andrew Tolley for helpful discussions. CB and PM are grateful to the organizers of the 2017 Royal Society Theo Murphy Scientific Meeting on ``Higgs Cosmology'', where this project was initiated.
\end{acknowledgments}


\appendix

\section{Linear order: Jordan frame}
\label{sec:linorderJ}

In order to study the dynamics of the Higgs and dilaton fields to linear order in fluctuations, we expand the system in Eqs.~\eqref{eq:system} and \eqref{eq:traceEinstein} in terms of deviations from the background field values of the scalar fields, $v_{\phi}=\braket{\phi}$ and $v_{\chi}=\braket{\chi}$, decomposing $\phi=v_{\phi}+\phi_1$ and $\chi=v_{\chi}+\chi_1$. The constant vevs $v_{\phi}$ and $v_{\chi}$ lie at one of the global minima of the potential $U(\phi,\chi)$ and are related via $v_{\phi}=\sqrt{\beta/\lambda}\,v_{\chi}$.

We imagine that the linear perturbations $\phi_1$ and $\chi_1$ are sourced by fermion condensates $\braket{\bar{\psi}\psi}$ and $\braket{\bar{N}N}$. Working on a background Minkowski spacetime, these fermion condensates source a scalar curvature $\mathcal{R}_1$, and we have
\begin{subequations}
\begin{gather}
\label{eq:linearsystem1}
\partial^2\phi_1\:+\:\xi_{\phi}\,\mathcal{R}_1\,v_{\phi}\:-\:m_{\phi}^2\,\phi_1\:+\:\frac{v_\chi}{v_{\phi}}\,m_{\chi}^2\,\chi_1\:-\:\braket{\bar{\psi}\psi}\ =\ 0\;,\\
\label{eq:linearsystem2}
\partial^2\chi_1\:+\:\xi_{\chi}\,\mathcal{R}_1\,v_{\chi}\:-\:m_{\chi}^2\,\chi_1\:+\:\frac{v_{\phi}}{v_{\chi}}\,m_{\phi}^2\,\phi_1\:-\:\braket{\bar{N}N}\ =\ 0\;,\\
\label{eq:constraintexp}
\partial^2\,\chi_1\ =\ -\,\frac{g_{\phi}}{g_{\chi}}\,\frac{v_{\phi}}{v_{\chi}}\,\partial^2\,\phi_1\;,
\end{gather}
\end{subequations}
where
\begin{equation}
m_{\phi}^2\ \equiv\ 2\lambda v_{\phi}^2\;,\qquad m_{\chi}^2\ \equiv\ 2\beta v_{\phi}^2\;.
\end{equation}
Here, we have made use of the fact that the background scalar curvature is zero, and the metric perturbations are sourced at linear order by the gradient energies of $\phi_1$ and $\chi_1$ and the fermion mass terms:
\begin{align}
\mathcal{R}_1\ &=\ -\:\frac{1}{M_{\rm Pl}^2}\Big(v_{\phi}\, \partial^2\phi_1+v_{\chi}\, \partial^2\chi_1\Big)\:+\:\frac{m_{\psi}}{M_{\rm Pl}^2}\,\braket{\bar{\psi}\psi}\:+\:\frac{m_N}{M_{\rm Pl}^2}\,\braket{\bar{N}N}\nonumber\\ &=\ \frac{1}{M_{\rm Pl}^2}\bigg(\frac{g_{\phi}}{g_{\chi}}-1\bigg)v_{\phi}\,\partial^2\phi_1\:+\:\frac{m_{\psi}}{M_{\rm Pl}^2}\,\braket{\bar{\psi}\psi}\:+\:\frac{m_N}{M_{\rm Pl}^2}\,\braket{\bar{N}N}\nonumber\\&=\ \frac{1}{M_{\rm Pl}^2}\bigg(\frac{g_{\chi}}{g_{\phi}}-1\bigg)v_{\chi}\,\partial^2\chi_1\:+\:\frac{m_{\psi}}{M_{\rm Pl}^2}\,\braket{\bar{\psi}\psi}\:+\:\frac{m_N}{M_{\rm Pl}^2}\,\braket{\bar{N}N}\;,
\end{align}
where $M_{\rm Pl}^2\equiv \xi_{\phi}v_{\phi}^2+\xi_{\chi}v_{\chi}^2$, and $m_{\psi}\equiv v_{\phi}$ and $m_{N}\equiv v_{\chi}$ (not to be confused with the nucleon mass appearing in Subsec.~\ref{sec:hadron}) are the fermion masses. Notice that no metric perturbations are sourced by the gradients of the scalar fields in the limit $g_{\phi}/g_{\chi}=1$.

Substituting for $\mathcal{R}_1$ in Eqs.~\eqref{eq:linearsystem1} and \eqref{eq:linearsystem2}, we can now write the scalar equations of motion	 in the form
\begin{subequations}
\begin{align}
\label{eq:deltaheom}
Z_{\phi}\partial^2\phi_1\:-\:m_{\phi}^2\,\phi_1\:+\:\frac{v_\chi}{v_{\phi}}\,m_{\chi}^2\,\chi_1\ &=\ +\:\frac{\xi_{\chi}v_{\chi}^2}{M_{\rm Pl}^2}\,\frac{m_{\psi}}{v_{\phi}}\,\braket{\bar{\psi}\psi}\:-\:\frac{\xi_{\phi} v_{\phi}^2}{M_{\rm Pl}^2}\,\frac{m_N}{v_{\phi}}\,\braket{\bar{N}N}\;,\\
Z_{\chi}\partial^2\chi_1\:-\:m_{\chi}^2\,\chi_1\:+\:\frac{v_{\phi}}{v_{\chi}}\,m_{\phi}^2\,\phi_1\ &=\ -\:\frac{\xi_\chi v_\chi^2}{M_{\rm Pl}^2}\,\frac{m_{\psi}}{v_{\chi}}\,\braket{\bar{\psi}\psi}\:+\:\frac{\xi_{\phi} v_{\phi}^2}{M_{\rm Pl}^2}\,\frac{m_N}{v_{\chi}}\,\braket{\bar{N}N}\;,
\end{align}
\end{subequations}
where we have defined
\begin{equation}
Z_{\phi}\ \equiv\ 1\:+\:\frac{\xi_{\phi}v_{\phi}^2}{M_{\rm Pl}^2}\bigg(\frac{g_{\phi}}{g_{\chi}}-1\bigg)\;,\qquad Z_{\chi}\ \equiv\ 1\:+\:\frac{\xi_{\chi}v_{\chi}^2}{M_{\rm Pl}^2}\bigg(\frac{g_{\chi}}{g_{\phi}}-1\bigg)\;.
\end{equation}
We see immediately that the massless mode (the dilaton):
\begin{equation}
\label{eq:dilaton1}
\sigma_1\ \propto\ Z_{\phi}\,\phi_1\:+\:\frac{v_{\chi}}{v_{\phi}}\,Z_{\chi}\,\chi_1
\end{equation}
is not sourced by either of the fermion fields. Remarkably, we also see that, if $\xi_{\phi}=0$, $\chi_1$ does not actually couple to $\bar{N}N$, i.e.~the gravitational backreaction exactly cancels the Yukawa coupling $\bar{N}\chi N$. Conversely, if $\xi_\chi = 0$, $\phi_1$ does not couple to $\bar{\psi}\psi$. Notice that non-minimal couplings can modify the Yukawa couplings (see also Ref.~\cite{Bezrukov:2014ipa}), and it is therefore not correct to ignore the non-minimal coupling to the scalar curvature in vacuum, where one might naively assume it is irrelevant, since $\mathcal{R}=0$ at the background level.


\section{Linear order: Einstein frame}
\label{sec:linearE}

We now repeat the linear-order analysis in the Einstein frame to illustrate the equivalence with the Jordan-frame analysis of App.~\ref{sec:linorderJ}. We map to the Einstein frame via the following Weyl rescaling of the metric
\begin{equation}
g_{\mu\nu}\ =\ \frac{M_{\rm Pl}^2}{F(\phi,\chi)}\,\tilde{g}_{\mu\nu}\;,
\end{equation}
where $F(\phi,\chi)$ is defined in Eq.~\eqref{eq:Fdef}.  The Einstein-frame action takes the form
\begin{align}\label{Einstein-Frame-Action}
S\ =\ \int\!{\rm d}^4x\;\sqrt{-\,\tilde{g}}\;\bigg[\frac{M_{\rm Pl}^2}{2}\,\tilde{\mathcal{R}}\:+\:\tilde{\mathcal{L}}_{\phi,\chi}\:+\:\frac{M_{\rm Pl}^4}{F^2(\phi,\chi)}\,\mathcal{L}_{\psi,N}\bigg]\;,
\end{align}
where
\begin{align}
\tilde{\mathcal{L}}_{\phi,\chi}\ &=\ -\:\frac{1}{2}\,\frac{M_{\rm Pl}^2}{F(\phi,\chi)}\,\tilde{g}^{\mu\nu}\,\partial_{\mu}\phi\,\partial_{\nu}\phi\:-\:\frac{1}{2}\,\frac{M_{\rm Pl}^2}{F(\phi,\chi)}\,\tilde{g}^{\mu\nu}\,\partial_{\mu}\chi\,\partial_{\nu}\chi\nonumber\\&\qquad-\:\frac{3}{4}\,\tilde{g}^{\mu\nu}\,M_{\rm Pl}^2\,\frac{\partial_{\mu}F(\phi,\chi)}{F(\phi,\chi)}\,\frac{\partial_{\nu}F(\phi,\chi)}{F(\phi,\chi)}\:-\:\frac{M_{\rm Pl}^4}{F^2(\phi,\chi)}\,U(\phi,\chi)\;.
\end{align}
Making the field redefinitions
\begin{equation}
\label{eq:fieldredefs}
\tilde{\psi}\ =\ \frac{M_{\rm Pl}^{3/2}\psi}{F^{3/4}(\phi,\chi)}\;,\qquad \tilde{N}\ =\ \frac{M_{\rm Pl}^{3/2}N}{F^{3/4}(\phi,\chi)}\;,
\end{equation}
the matter Lagrangian can be written as
\begin{equation}
\frac{M_{\rm Pl}^4}{F^2(\phi,\chi)}\,\mathcal{L}_{\psi,N}\ =\ -\:\bar{\tilde{\psi}}i\overset{\leftrightarrow}{\tilde{\slashed{\partial}}}\tilde{\psi}\:-\:\bar{\tilde{N}}i\overset{\leftrightarrow}{\tilde{\slashed{\partial}}}\tilde{N}\:-\:\frac{M_{\rm Pl}}{F^{1/2}(\phi,\chi)}\Big(\bar{\tilde{\psi}}\phi\tilde{\psi}\:+\:\bar{\tilde{N}}\chi \tilde{N}\:+\:\bar{\tilde{\psi}}\phi \tilde{N}\:+\:\bar{\tilde{N}}\phi \tilde{\psi}\Big)\;.
\end{equation}

At linear order, we find the equations of motion
\begin{subequations}
\begin{align}
\label{eq:Eins1}
&\bigg[1\:+\:6\,\frac{\xi_{\phi}^2v_{\phi}^2}{M_{\rm Pl}^2}\bigg]\,\partial^2\phi_1\:+\:6\,\frac{\xi_{\phi}\xi_{\chi}v_{\phi}v_{\chi}}{M_{\rm Pl}^2}\,\partial^2\chi_1\:-\:m_{\phi}^2\,\phi_1\:+\:\frac{v_{\chi}}{v_{\phi}}\,m_{\chi}^2\,\chi_1\nonumber\\&\qquad =\ +\:\frac{\xi_{\chi}v_{\chi}^2}{M_{\rm Pl}^2}\,\frac{m_{\psi}}{v_{\phi}}\,\braket{\bar{\tilde{\psi}}\tilde{\psi}}\:-\:\frac{\xi_{\phi}v_{\phi}^2}{M_{\rm Pl}^2}\,\frac{m_{N}}{v_{\phi}}\,\braket{\bar{\tilde{N}}\tilde{N}}\;,\\
\label{eq:Eins2}
&\bigg[1\:+\:6\,\frac{\xi_{\chi}^2v_{\chi}^2}{M_{\rm Pl}^2}\bigg]\,\partial^2\chi_1\:+\:6\,\frac{\xi_{\phi}\xi_{\chi}v_{\phi}v_{\chi}}{M_{\rm Pl}^2}\,\partial^2\phi_1\:-\:m_{\chi}^2\,\chi_1\:+\:\frac{v_{\phi}}{v_{\chi}}\,m_{\phi}^2\,\phi_1\nonumber\\&\qquad =\ -\:\frac{\xi_{\chi}v_{\chi}^2}{M_{\rm Pl}^2}\,\frac{m_{\psi}}{v_{\chi}}\,\braket{\bar{\tilde{\psi}}\tilde{\psi}}\:+\:\frac{\xi_{\phi}v_{\phi}^2}{M_{\rm Pl}^2}\,\frac{m_{N}}{v_{\chi}}\,\braket{\bar{\tilde{N}}\tilde{N}}\;.
\end{align}
\end{subequations}
Notice that, since both $U(\phi,\chi)$ and its first derivatives vanish when evaluated at $\phi=v_{\phi}$ and $\chi=v_{\chi}$, the second variation of the potential yields the same mass terms as in the Jordan frame at linear order. Moreover, since \smash{$\braket{\bar{\tilde{\psi}}\tilde{\psi}}=\braket{\bar{\psi}\psi}$} and \smash{$\braket{\bar{\tilde{N}}\tilde{N}}=\braket{\bar{N}N}$} at this order, the source terms are also the same. By adding to and subtracting from Eq.~\eqref{eq:Eins1} $v_{\chi}/v_{\phi}$ times Eq.~\eqref{eq:Eins2}, we can quickly confirm that we recover precisely the results obtained in the Jordan frame in App.~\ref{sec:linorderJ}.

\end{document}